\documentclass[letterpaper,english,notitlepage,twocolumn, superscriptaddress]{revtex4-1}
\usepackage[T1]{fontenc}
\usepackage[latin9]{inputenc}
\usepackage{amsmath}
\usepackage{amssymb}
\usepackage{graphicx}

\makeatletter

\pdfpageheight\paperheight
\pdfpagewidth\paperwidth

\makeatother

\usepackage{babel}
\begin{document}
\title{Optimal control theory for maximum output power of Brownian heat engines}
\author{Jin-Fu Chen}
\thanks{chenjinfu@pku.edu.cn}
\affiliation{School of Physics, Peking University, Beijing, 100871, China}
\author{H. T. Quan}
\thanks{htquan@pku.edu.cn}
\affiliation{School of Physics, Peking University, Beijing, 100871, China}
\affiliation{Collaborative Innovation Center of Quantum Matter, Beijing 100871,
China}
\affiliation{Frontiers Science Center for Nano-optoelectronics, Peking University,
Beijing, 100871, China}
\date{\today}
\begin{abstract}
The pursuit of achieving the maximum output power in microscopic heat
engines has gained increasing attention in the field of stochastic
thermodynamics. We employ the optimal control theory to study Brownian
heat engines and determine the optimal heat-engine cycles in generic
damped situation, which were previously known only in the overdamped
and the underdamped limits. These optimal cycles include two isothermal
processes, two adiabatic processes, and an extra isochoric relaxation
process at the high stiffness constraint. Our results determine the
maximum output power under realistic control constraints, and also
bridge the gap of the optimal cycles between the overdamped and the
underdamped limits. Hence, we solve an outstanding problem in the
studies of heat engines by employing the optimal control theory to
stochastic thermodynamics. These findings bring valuable insights
for the design of high-performance Brownian heat engines in experimental
setups.
\end{abstract}
\maketitle
\textit{Introduction. }The output power is an important index to characterize
the performance of real heat engines. To achieve efficient energy
conversion at the nanoscale, Brownian heat engines were invented to
extract useful work by harnessing the fluctuation and dissipation
of a single Brownian particle \citep{Blickle2011,QuintoSu2014,Martinez2015,Krishnamurthy2016,Martinez2017,Holubec2021}.
The output power of these microscopic heat engines is significantly
influenced by the coupling strength between the Brownian particle
and the heat bath (characterized by the friction coefficient) and
the control protocol of heat-engine cycles. However, what are the
optimal friction coefficient and the optimal control protocol to achieve
the maximum output power remain outstanding unsolved problems in the
studies of heat engines. For Brownian heat engines, the previous optimization
was limited to either the overdamped limit (large friction coefficient)
\citep{Schmiedl2007,schmiedl2007efficiency,Zulkowski2015,Plata2019,Proesmans2020,Ye2022,Xu2022,Abiuso2022}
or the underdamped limit (small friction coefficient) \citep{Dechant2017,Chen2022}.
In the underdamped limit, the efficiency at the maximum output power
agrees with Curzon and Ahborn's prediction \citep{curzon1975efficiency}.
The output power diminishes in both the overdamped and the underdamped
limits. It is expected that larger output power can be achieved with
intermediate friction coefficient. Nevertheless, all the previous
methods fail in this regime \citep{Dechant2017}.

Recently, the thermodynamic geometry \citep{Salamon1983,Crooks2007,Sivak2012,Zulkowski2012,Scandi2019,Blaber2023}
has been applied to optimize control protocols of thermodynamic processes
in the slow-driving regime \citep{Cavina2017,Brandner2020,Chen2021,Abiuso2020,Li2022,Chen2022a,Frim2022,Wang2023,Chen2023a}.
However, these slow-driving protocols are not necessarily optimal
to achieve the maximum output power since the optimal heat-engine
cycles may operate far beyond the slow-driving regime. By contrast,
a more elaborate optimization method is to employ the optimal control
theory \citep{pontryagin1962}, which has been extensively employed
in different subfields of physics, such as, optimal controls of closed
and open quantum systems \citep{Peirce1988,Caneva2009,Lloyd2014,Li2017,Cavina2018,Boscain2021}
and quantum algorithms \citep{Yang2017,Brady2021}. The optimal control
theory has also been employed to optimize ideal-gas heat-engine cycles
in the framework of phenomenological finite-time thermodynamics \citep{Rubin1979a,Rubin1980,Mozurkewich1982,Andresen1984a,Hoffmann1985}.
Nevertheless, these models are based on the phenomenological assumption
of Newton\textquoteright s cooling law and lack the equation of motion.
Hence, their results were still limited to the weak-coupling regime.
Stochastic thermodynamics, on the other hand, is a framework of nonequilibrium
thermodynamics characterizing irreversibility and fluctuations based
on the equation of motion. It is valid for arbitrary friction coefficient,
and it brings more insights such as fluctuation theorems \citep{Jarzynski1997,seifert2012stochastic}
and thermodynamic uncertainty relation \citep{barato2015thermodynamic,Horowitz2017}
than the phenomenological finite-time thermodynamics. However, the
optimal control theory has not been applied to stochastic thermodynamics
previously (see Ref. \citep{MuratoreGinanneschi2014}).

In this Letter, we employ the optimal control theory to investigate
the maximum output power of Brownian heat engines in the framework
of stochastic thermodynamics. Our method of determining the optimal
control protocol is valid for arbitrary friction coefficient. Under
given control constraints of the stiffness and the bath temperature,
we find a family of cycles with conserved pseudo-Hamiltonian, which
is introduced by the Pontryagin maximum principle \citep{pontryagin1962}.
The optimal cycle is the one with the largest pseudo-Hamiltonian promising
the maximum output power. Based on this microscopic model, we also
bridge the gap of the optimal cycles between the overdamped \citep{Schmiedl2007,schmiedl2007efficiency}
and the underdamped limits \citep{Dechant2017,Chen2022}. Thus, by
employing the optimal control theory, we solve the outstanding problems
of determining the optimal friction coefficient and the optimal control
protocol to achieve the maximum output power of Brownian heat engines.
Our results indicate that the most powerful heat engines operate at
an intermediate friction coefficient, neither too small (underdamped
limit) nor too large (overdamped limit). Our optimal cycles guide
the design of Brownian heat engines in experimental setups \citep{Martinez2015,Hoang2018,Ferrer2021}.

\textit{Setup and optimal isothermal processes.} We consider a microscopic
heat engine with a single Brownian particle as the working substance
\citep{Blickle2011,Martinez2015,schmiedl2007efficiency,Chen2022}.
The Hamiltonian of the system is

\begin{equation}
H=\frac{p^{2}}{2}+\frac{1}{2}\lambda x^{2},
\end{equation}
where the mass of the particle is set unity, and the work parameter
$\lambda$ is the stiffness of the harmonic potential. The system
is coupled to a heat bath at temperature $T_{b}$. The stiffness $\lambda$
and the bath temperature $T_{b}$ can be varied by an external agent
to construct heat-engine cycles. The evolution of the distribution
$\rho(x,p,t)$ of position and momentum is described by the Fokker-Planck
equation \citep{Kramers1940}

\begin{equation}
\frac{\partial\rho}{\partial t}=-\frac{\partial}{\partial x}(p\rho)+\frac{\partial}{\partial p}(\lambda x\rho+\kappa p\rho+\kappa T_{b}\frac{\partial\rho}{\partial p}),
\end{equation}
with the friction coefficient $\kappa$. The distribution remains
Gaussian during the evolution when the system evolves from an initial
Gaussian state. We characterize the state of the system with the covariances
of position and momentum $\Gamma=(\Gamma_{xx},\Gamma_{xp},\Gamma_{pp})^{T}$
with $\Gamma_{xx}=\left\langle x^{2}\right\rangle $, $\Gamma_{xp}=\left\langle xp\right\rangle $,
and $\Gamma_{pp}=\left\langle p^{2}\right\rangle $, satisfying \citep{Rezek2006,Chen2023}

\begin{align}
\dot{\Gamma} & =L\Gamma+f,\label{eq:first-order_eq}\\
L & =\left(\begin{array}{ccc}
0 & 2 & 0\\
-\lambda & -\kappa & 1\\
0 & -2\lambda & -2\kappa
\end{array}\right),\label{eq:L_mat}
\end{align}
with $f=(0,0,2\kappa T_{b})^{T}.$ The real parts of $L$'s eigenvalues
$l_{1}=-\kappa+\sqrt{\kappa^{2}-4\lambda}$, $l_{2}=-\kappa$, and
$l_{3}=-\kappa-\sqrt{\kappa^{2}-4\lambda}$ determine the relaxation
rates of covariances. The relaxation rate of position is determined
by the minimal real part of these eigenvalues, i.e., $\Delta=-\mathrm{Re}(l_{1})$.
The work performed on the system is $\dot{W}=\dot{\lambda}\Gamma_{xx}/2$.
By substituting $\Gamma_{xp}=\dot{\Gamma}_{xx}/2$ and $\Gamma_{pp}=(\kappa\dot{\Gamma}_{xx}+2\lambda\Gamma_{xx}+\ddot{\Gamma}_{xx})/2$
into Eq. (\ref{eq:first-order_eq}), we obtain a high-order differential
equation of $\Gamma_{xx}$ as

\begin{equation}
\dddot{\Gamma}_{xx}+3\kappa\ddot{\Gamma}_{xx}+(2\kappa^{2}+4\lambda)\dot{\Gamma}_{xx}=4\kappa T_{b}-(2\dot{\lambda}+4\kappa\lambda)\Gamma_{xx}.\label{eq:Gamma_xxdddot}
\end{equation}
It is challenging to solve the optimal control problem of minimizing
the performed work based on the exact equation (\ref{eq:first-order_eq})
of motion {[}or Eq. (\ref{eq:Gamma_xxdddot}){]}, since the system
is not fully controllable \citep{pontryagin1962}: there are three
state variables $\Gamma_{xx}$, $\Gamma_{xp}$ and $\Gamma_{pp}$
but only two parameters $\lambda$ and $T_{b}$ can be varied.

To simplify the optimal control problem, we focus on relatively slow
control $\dot{\lambda}/\lambda<\Delta$ with slowly varying $\dot{\Gamma}_{xx}$
such that the higher-order derivatives in Eq. (\ref{eq:Gamma_xxdddot})
can be neglected. Such approximation renders the system controllable
\citep{pontryagin1962}, and Eq. (\ref{eq:Gamma_xxdddot}) is simplified
into
\begin{equation}
\dot{\Gamma}_{xx}=-\frac{2\kappa\lambda}{\kappa^{2}+2\lambda}(\Gamma_{xx}-\frac{T_{b}}{\lambda})-\frac{\dot{\lambda}}{\kappa^{2}+2\lambda}\Gamma_{xx},\label{eq:dotgammaxxapproximate}
\end{equation}
which is rewritten in the form of the potential energy $V\equiv\lambda\Gamma_{xx}/2$
as

\begin{equation}
\dot{V}=\frac{(\kappa^{2}+\lambda)\dot{\lambda}}{(\kappa^{2}+2\lambda)\lambda}V-\frac{\kappa\lambda(2V-T_{b})}{(\kappa^{2}+2\lambda)}.\label{eq:dot_V}
\end{equation}
We will use the approximate equation (\ref{eq:dot_V}) of motion to
solve the optimal control protocol of varying $\lambda(t)$, and propagate
the exact equation (\ref{eq:first-order_eq}) of motion with the solved
protocol. The optimal cycles, constructed by the optimal control protocols
and the switchings, indeed leads to large output power for the intermediate
friction coefficient, where the previous protocols of heat-engine
cycles \citep{Schmiedl2007,schmiedl2007efficiency,Dechant2017,Chen2022}
in the underdamped and overdamped limits fail. We will validate the
approximate equation (\ref{eq:dotgammaxxapproximate}) of motion with
different friction coefficients in the domain of its applicability
$\dot{\lambda}/\lambda<\Delta$.

We now employ the optimal control theory to the control of the stiffness
and the bath temperature. The potential energy $V$ and the stiffness
$\lambda$ are state variables in the control problem. The control
parameters are the rate of stiffness change $\alpha=\dot{\lambda}/\lambda$
and the bath temperature $T_{b}$. The control constraints are $\lambda_{L}\leq\lambda\leq\lambda_{H}$
and $T_{L}\le T_{b}\le T_{H}$. The output work of a cycle is $W_{\mathrm{out}}=-\int_{0}^{\tau}\alpha Vdt$,
and is the goal of the optimization. The pseudo-Hamiltonian of the
optimal control theory for this system is constructed as \citep{SupplementaryMaterials}

\begin{align}
\mathcal{H} & =-\alpha V+\psi_{V}\frac{(\kappa^{2}+\lambda)\alpha V-\kappa\lambda(2V-T_{b})}{\kappa^{2}+2\lambda}+\psi_{\lambda}\alpha\lambda,
\end{align}
where $\psi_{V}$ and $\psi_{\lambda}$ are costate variables \citep{pontryagin1962,Rubin1979a}.
The pseudo-Hamiltonian $\mathcal{H}$ is introduced to derive the
optimal control protocol, and should be distinguished from the Hamiltonian
$H$ of the system. The evolution of the state variables is governed
by $\dot{V}=\partial\mathcal{H}/\partial\psi_{V}$ {[}Eq. (\ref{eq:dot_V}){]}
and $\dot{\lambda}=\partial\mathcal{H}/\partial\psi_{\lambda}=\alpha\lambda$,
while the evolution of the costate variables is governed by
\begin{align}
\dot{\psi}_{V} & =-\frac{\partial\mathcal{H}}{\partial V}=\alpha-\psi_{V}\frac{\alpha\left(\kappa^{2}+\lambda\right)-2\kappa\lambda}{\kappa^{2}+2\lambda},\label{eq:dot_psi_V}\\
\dot{\psi}_{\lambda} & =-\frac{\partial\mathcal{H}}{\partial\lambda}=\psi_{V}\frac{\kappa^{2}\left(V(\alpha+2\kappa)-\kappa T_{b}\right)}{\left(\kappa^{2}+2\lambda\right)^{2}}-\alpha\psi_{\lambda}.\label{eq:dot_psi_lambda}
\end{align}

The control parameters $\alpha$ and $T_{b}$ are chosen to maximize
$\mathcal{H}$ under the control constraints. The control protocol
of the bath temperature $T_{b}$ is determined by

\begin{equation}
\frac{\partial\mathcal{H}}{\partial T_{b}}=\frac{\kappa\lambda\psi_{V}}{\kappa^{2}+2\lambda},
\end{equation}
which corresponds to a bang-bang control $T_{b}=T_{H}$ for $\psi_{V}>0$
and $T_{b}=T_{L}$ for $\psi_{V}<0$. The control protocol of the
stiffness change rate $\alpha$ is determined by

\begin{equation}
\frac{\partial\mathcal{H}}{\partial\alpha}=\lambda\psi_{\lambda}+\frac{V\left(\kappa^{2}+\lambda\right)\psi_{V}}{\kappa^{2}+2\lambda}-V.\label{eq:partialHpartialalpha}
\end{equation}
Due to the fact that the right hand side of Eq. (\ref{eq:partialHpartialalpha})
does not depend on $\alpha$, we first consider the singular control
with $\partial\mathcal{H}/\partial\alpha=0$ \citep{pontryagin1962}.
The costate variables and the rate of stiffness change are solved
as

\begin{align}
\psi_{V} & =\frac{\left(\kappa^{2}+2\lambda\right)\left(T_{b}-2V\right)}{\lambda T_{b}+2\kappa^{2}V},\label{eq:psi_V}\\
\psi_{\lambda} & =\frac{V\left[2V\left(2\kappa^{2}+\lambda\right)-\kappa^{2}T_{b}\right]}{\lambda\left(\lambda T_{b}+2\kappa^{2}V\right)},\label{eq:psi_lambda}\\
\frac{\dot{\lambda}}{\lambda} & =\frac{\kappa\lambda(2V-T_{b})\left[2\kappa^{2}V+(\kappa^{2}+2\lambda)T_{b}\right]}{V\left[4\kappa^{4}V+6\kappa^{2}\lambda V+(\kappa^{2}+2\lambda)\lambda T_{b}\right]},\label{eq:dot_Lambda_approximate_dynamics}
\end{align}
where Eq. (\ref{eq:dot_Lambda_approximate_dynamics}) gives the optimal
control protocols of isothermal processes.

We eliminate the costate variables with Eqs. (\ref{eq:psi_V}) and
(\ref{eq:psi_lambda}), and obtain the pseudo-Hamiltonian

\begin{equation}
\mathcal{H}=\frac{\kappa\lambda\left(T_{b}-2V\right){}^{2}}{\lambda T_{b}+2\kappa^{2}V},\label{eq:pseudo_h_general}
\end{equation}
whose magnitude encodes the information of the velocity of the control.
In quasistatic isothermal processes, the system is in equilibrium
with the heat bath, i.e., $V=T_{b}/2$, and thus $\mathcal{H}=0$.
In nonquasistatic isothermal processes, we can solve $V$ of the optimal
control from the conservation of $\mathcal{H}$ as
\begin{equation}
V=\frac{T_{b}}{2}+\frac{\kappa\mathcal{H}}{4\lambda}[1-\epsilon\sqrt{1+\frac{4\lambda T_{b}}{\kappa\mathcal{H}}(1+\frac{\lambda}{\kappa^{2}})}],
\end{equation}
where $\epsilon=+1$ for the expansion process and $-1$ for the compression
process.

The control protocol can be significantly simplified in the two limits.
In the overdamped limit $\kappa^{2}\gg\lambda$, the pseudo-Hamiltonian
Eq. (\ref{eq:pseudo_h_general}) becomes $\mathcal{H}=\lambda(T_{b}-2V)^{2}/(2\kappa V)$,
and the control protocol of the stiffness and the evolution of the
potential energy are

\begin{align}
\lambda(t) & =\frac{\lambda_{0}[1-\mathcal{H}t/(2V_{0})]}{\left(1+\epsilon\sqrt{\lambda_{0}\mathcal{H}/(2\kappa V_{0})}t\right)^{2}},\label{eq:Lambdat_overdamped}\\
V(t) & =V_{0}-\mathcal{H}t/2.\label{eq:Vt_overdamped}
\end{align}
The above protocol agrees with that obtained in Ref. \citep{Schmiedl2007}.
In the underdamped limit $\kappa^{2}\ll\lambda$, the pseudo-Hamiltonian
Eq. (\ref{eq:pseudo_h_general}) becomes $\mathcal{H}=\kappa(T_{b}-2V)^{2}/T_{b}$.
The potential energy $V$ of the optimal control thus remains constant
in nonquasistatic isothermal processes

\begin{equation}
V(t)=\frac{T_{b}}{2}-\frac{\epsilon}{2}\sqrt{\frac{\mathcal{H}T_{b}}{\kappa}},\label{eq:Vt_underdamped}
\end{equation}
and the control protocol is the exponential protocol

\begin{equation}
\frac{\dot{\lambda}}{\lambda}=\frac{2\kappa}{1-\epsilon\sqrt{\kappa T_{b}/\mathcal{H}}}=\mathrm{const},\label{eq:Lambdat_underdamped}
\end{equation}
which agrees with that obtained in Refs. \citep{Gong2016,salazar2020work}.
\begin{figure}
\includegraphics[width=8cm]{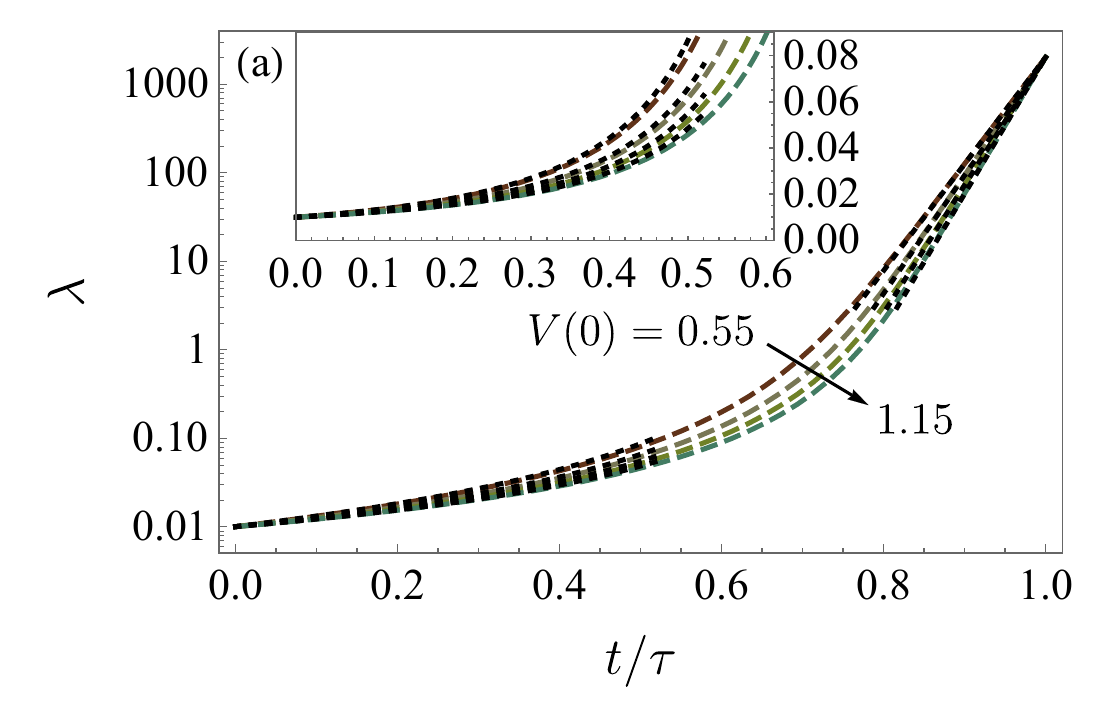}

\includegraphics[width=8cm]{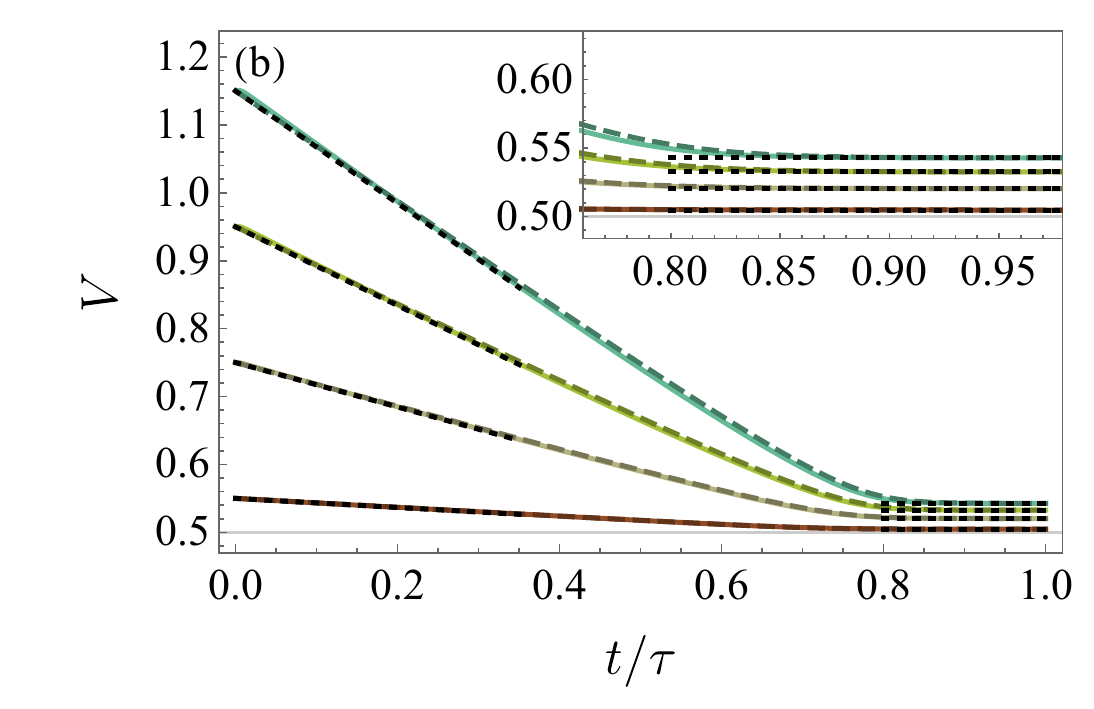}

\caption{The optimal control of isothermal processes in the generic damped
situation (arbitrary friction coefficient). The parameters are set
as $T_{b}=1$ and $\kappa=1$. (a) The optimal control protocols with
different control speeds characterized by $V(0)=0.55$, $0.75$, $0.95$,
and $1.15$. We plot all the protocols (dashed curves) with a rescaled
time $t/\tau$ and the initial and final stiffness $\lambda(0)=0.01$
and $\lambda(\tau)=2000$. (b) The evolution of the potential energy
$V$ during the isothermal processes. The solid curves show the results
of the exact equation (\ref{eq:first-order_eq}) of motion with the
protocols in Fig. \ref{fig:The-protocols-and-evolution}(a). In both
subfigures, the dotted lines show the results in the overdamped and
the underdamped limits, and the dashed curves represent the approximate
results obtained from Eqs. (\ref{eq:dot_V}) and (\ref{eq:dot_Lambda_approximate_dynamics}).
\label{fig:The-protocols-and-evolution}}
\end{figure}

It is desirable to check the validity of the approximation used in
Eq. (\ref{eq:dotgammaxxapproximate}). After solving the protocols
from Eqs. (\ref{eq:dot_V}) and (\ref{eq:dot_Lambda_approximate_dynamics}),
we can utilize these protocols to propagate the exact equation (\ref{eq:first-order_eq})
of motion. Figure \ref{fig:The-protocols-and-evolution}(a) shows
the control protocols of the stiffness $\lambda(t)$ (dashed curves)
with a rescaled time $t/\tau$ for the isothermal compression processes.
The black dotted curves represent the optimal control protocols in
the overdamped limit \citep{Schmiedl2007,schmiedl2007efficiency}
and the underdamped limit \citep{Dechant2017,Chen2022}, agreeing
with our protocols at the initial and final stages, respectively.
The results of the evolution of potential energy are compared in Fig.
\ref{fig:The-protocols-and-evolution} (b). We set the bath temperature
$T_{b}=1$, the friction coefficient $\kappa=1$, and the initial
and final stiffness $\lambda(0)=0.01$ and $\lambda(\tau)=2000$,
and choose the initial potential energy as $V(0)=0.55$, $0.75$,
$0.95$, and $1.15$. With larger $V(0)$, the isothermal compression
is carried out more rapidly. The dashed curves are solved based on
the approximate equation (\ref{eq:dot_V}) of motion, while the solid
curves show the exact evolution by propagating Eq. (\ref{eq:first-order_eq})
with the protocols in Fig. \ref{fig:The-protocols-and-evolution}(a).
The agreement of the solid curves and the dashed curves validates
the approximation used in Eq. (\ref{eq:dotgammaxxapproximate}), at
least in the domain of its applicability.

\textit{Construction of the optimal cycles.} We continue to construct
heat-engine cycles under the control constraints $\lambda_{L}\leq\lambda\leq\lambda_{H}$
and $T_{L}\le T_{b}\le T_{H}$. During the whole cycle, the pseudo-Hamiltonian
is conserved and we thus require two isothermal processes with identical
$\mathcal{H}$. For a given $\mathcal{H}$, we find two switchings
at the stiffness constraints $\lambda_{L}$ and $\lambda_{H}$ are
necessary: $T_{b}$ and $\lambda$ are changed simultaneously and
instantaneously \citep{SupplementaryMaterials}. Thus, $\mathcal{H}$
together with the control constraints determines the control protocol
of the cycle.

\begin{figure}
\includegraphics[width=8cm]{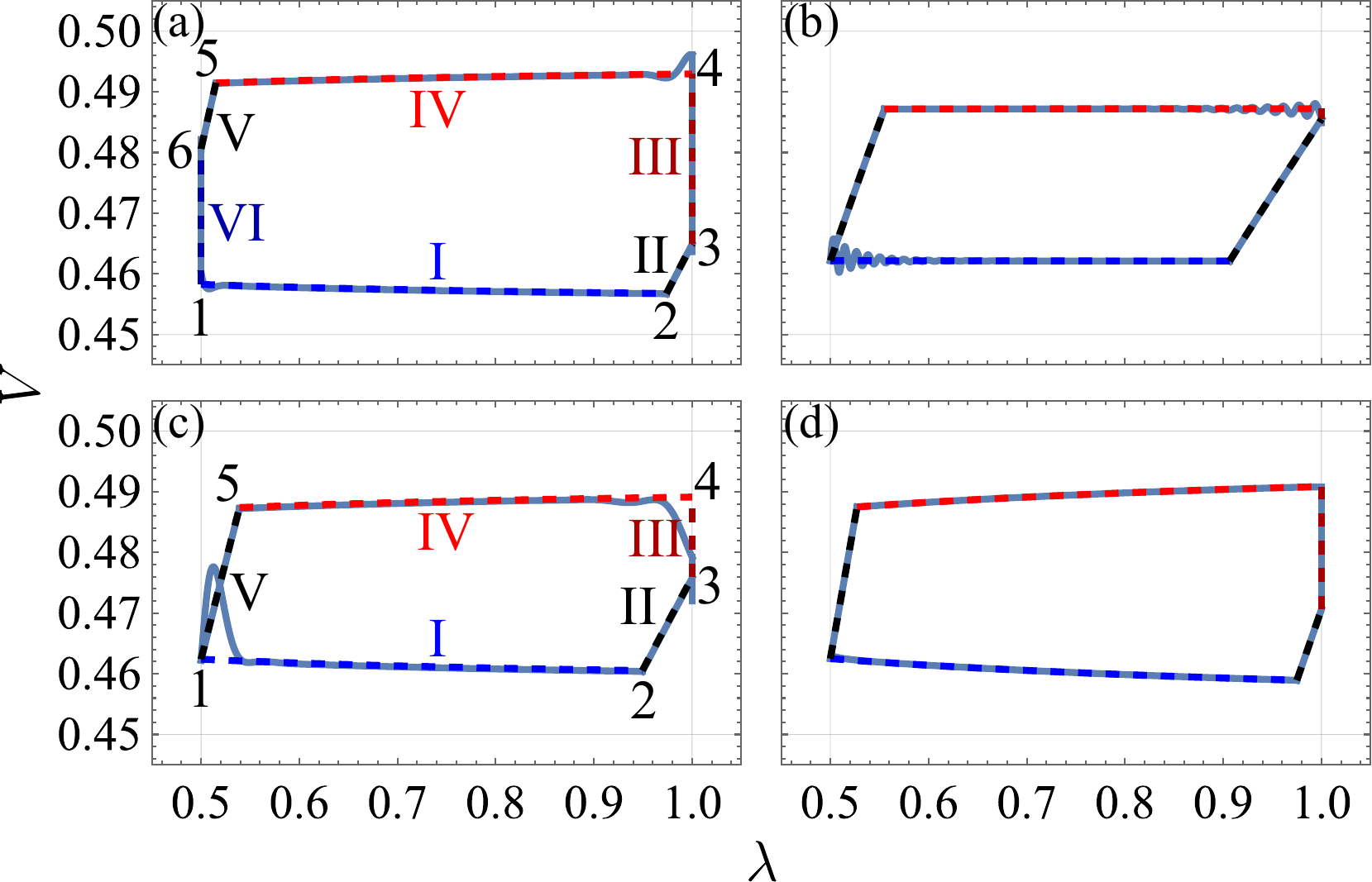}

\caption{Diagrams of Brownian heat-engine cycles. (a) $\kappa=1$ and $\mathcal{H}=10^{-4}$
(nonmaximum $\mathcal{H}$), (b) $\kappa=0.1$ (underdamped, maximum
$\mathcal{H}$), (c) $\kappa=0.83$ (optimal, maximum $\mathcal{H}$),
and (d) $\kappa=10$ (overdamped, maximum $\mathcal{H}$). The control
constraints are $T_{L}=0.9$, $T_{H}=1$, $\lambda_{L}=0.5$ and $\lambda_{H}=1$.
The solid and the dashed curves represent limit cycles according to
the exact equation (\ref{eq:first-order_eq}) and the approximate
equation (\ref{eq:dot_V}) of motion, respectively. \label{fig:cycle-diagram-and control scheme}}
\end{figure}

During the switchings, both $\lambda$ and $T_{b}$ are changed instantaneously,
i.e., $\alpha\gg\kappa$, and the evolution equations (\ref{eq:dot_V})
and (\ref{eq:dot_psi_V}) are reduced to 
\begin{align}
\dot{V} & =\frac{\kappa^{2}+\lambda}{\kappa^{2}+2\lambda}\alpha V,\\
\dot{\psi}_{V} & =\alpha\left[1-\frac{\left(\kappa^{2}+\lambda\right)}{\kappa^{2}+2\lambda}\psi_{V}\right].
\end{align}
The initial and final values during the switchings satisfy 
\begin{align}
\frac{V_{f}\sqrt{\kappa^{2}+2\lambda_{f}}}{\lambda_{i}} & =\frac{V_{i}\sqrt{2\lambda_{i}+\kappa^{2}}}{\lambda_{i}},\\
\frac{\lambda_{f}(\psi_{Vf}-2)-\kappa^{2}}{\sqrt{\kappa^{2}+2\lambda_{f}}} & =\frac{\lambda_{i}\left(\psi_{Vi}-2\right)-\kappa^{2}}{\sqrt{\kappa^{2}+2\lambda_{i}}}.
\end{align}
The two switchings II and V in a cycle are given by $\lambda_{2}\rightarrow\lambda_{H},V_{2}\rightarrow V_{3},\psi_{V2}\rightarrow\psi_{V3}$
and $\lambda_{5}\rightarrow\lambda_{L},V_{5}\rightarrow V_{6},\psi_{V5}\rightarrow\psi_{V6}$,
as shown in Fig. \ref{fig:cycle-diagram-and control scheme}(a). The
six boundary points and the six processes are denoted as $1,2,...,6$
and I, II, ..., VI, respectively.\textbf{ }The two switchings are
solved from the conservation of $\mathcal{H}$ \citep{SupplementaryMaterials}.
By increasing $\mathcal{H}$, point 6 will approach point 1 on the
cold isothermal curve. Since larger $\mathcal{H}$ leads to larger
output power, we consider the maximum $\mathcal{H}$ such that the
cycle still reaches $\lambda_{L}$. The output power $P$ of the maximum-$\mathcal{H}$
cycle is evaluated as

\begin{equation}
P=-\frac{W_{\mathrm{I}}+W_{\mathrm{II}}+W_{\mathrm{IV}}+W_{\mathrm{V}}}{\tau_{\mathrm{I}}+\tau_{\mathrm{III}}+\tau_{\mathrm{IV}}}.\label{eq:power_estimated}
\end{equation}
The expressions of the work and the operation time of each process
are given in \citep{SupplementaryMaterials}.

With the above protocol of a whole cycle, we can propagate the exact
equation (\ref{eq:first-order_eq}) of motion and obtain the exact
output power of the limit cycle. In accordance with Eq. (\ref{eq:power_estimated}),
we choose the connecting processes to be thermodynamic adiabatic (unitary)
processes that satisfy $\Gamma_{pp}^{f}\Gamma_{xx}^{f}=\Gamma_{pp}^{i}\Gamma_{xx}^{i}$.
In principle, such unitary evolution can be carried out in arbitrarily
short time by choosing a large Hamiltonian \citep{Mandelstam_and_Tamm,Margolus1998}.
The corresponding protocol to vary the Hamiltonian can be realized
by adopting the shortcut to adiabaticity \citep{Berry2009,GueryOdelin2019}.
After solving $\Gamma_{xx}$, $\Gamma_{xp}$, and $\Gamma_{pp}$ during
the exact evolution of the whole cycle, we use Eq. (\ref{eq:power_estimated})
to evaluate the exact output power.

We plot diagrams of heat-engine cycles in Fig. \ref{fig:cycle-diagram-and control scheme}
for a non-maximum-$\mathcal{H}$ cycle with (a) $\kappa=1$ and $\mathcal{H}=10^{-4}$,
and three maximum-$\mathcal{H}$ cycles with (b) $\kappa=0.1$ (underdamped),
(c) $\kappa=0.83$ (optimal), and (d) $\kappa=10$ (overdamped). In
the underdamped limit {[}Fig. \ref{fig:cycle-diagram-and control scheme}(b){]},
the optimal protocol of the cycle recovers that of the Curzon-Ahlborn
engine \citep{curzon1975efficiency,Chen2022}, and the effective temperature
of the Brownian particle remains constant ($V=\mathrm{const}$) during
the isothermal processes. For an intermediate friction coefficient
{[}Fig. \ref{fig:cycle-diagram-and control scheme}(c){]}, the exact
evolution (solid curves) deviates slightly from the approximate evolution
(dashed curves), and the exact output power is similar to the approximate
output power. In the overdamped limit {[}Fig. \ref{fig:cycle-diagram-and control scheme}(d){]},
the protocols of the two isothermal processes recover those in Refs.
\citep{schmiedl2007efficiency,Schmiedl2007}.

\textit{Maximum output power of Brownian heat engines.} For a given
friction coefficient $\kappa$, we evaluate the output power $P$
of maximum-$\mathcal{H}$ cycles. Furthermore, the friction coefficient
$\kappa$ can be adjusted to achieve the maximum output power. In
a generic damped situation (arbitrary friction coefficient), we solve
the maximum-$\mathcal{H}$ cycle numerically. Analytical results are
obtained in the overdamped limit and the underdamped limit.

In the overdamped limit $\kappa^{2}\gg\lambda_{H}$, the cooling rate
according to Eq. (\ref{eq:dot_V}) is equal to $2\lambda/\kappa$.
Thus, the output power is inversely proportional to the friction coefficient
$P_{\mathrm{over}}\propto1/\kappa$ \citep{SupplementaryMaterials}.
We can choose $\lambda_{L}$ as large as possible so that the maximum
output power $P_{\mathrm{over}}^{*}$ can be achieved by the sliced
cycles \citep{SupplementaryMaterials}

\begin{equation}
P_{\mathrm{over}}^{*}=\frac{\lambda_{H}T_{H}}{\kappa}\phi(\frac{T_{L}}{T_{H}}),\label{eq:maximum_power_over_sliced}
\end{equation}
where $\phi(\theta)$ is the positive root of the polynomial

\begin{equation}
\phi^{3}+(11-3\theta)\phi^{2}+\left(3\theta^{2}+14\theta-1\right)\phi-(\theta-1)^{2}\theta=0,\label{eq:the polynomial}
\end{equation}
with $\theta=T_{L}/T_{H}$. A detailed discussion of these sliced
cycles is left in \citep{SupplementaryMaterials}.

In the underdamped limit $\kappa^{2}\ll\lambda_{L}$, the maximum
output power is \citep{Dechant2017,Chen2022}

\begin{equation}
P_{\mathrm{under}}^{*}=\frac{\kappa T_{H}}{4}(1-\sqrt{\frac{T_{L}}{T_{H}}})^{2}.\label{eq:maximum_power_under}
\end{equation}
We left the derivations of Eqs. (\ref{eq:maximum_power_over_sliced})
and (\ref{eq:maximum_power_under}) in \citep{SupplementaryMaterials}.
This maximum output power also agrees with the results of ideal-gas
heat engines based on Newton\textquoteright s cooling law \citep{Rubin1979a,Rubin1980}.

\begin{figure}
\includegraphics[width=8cm]{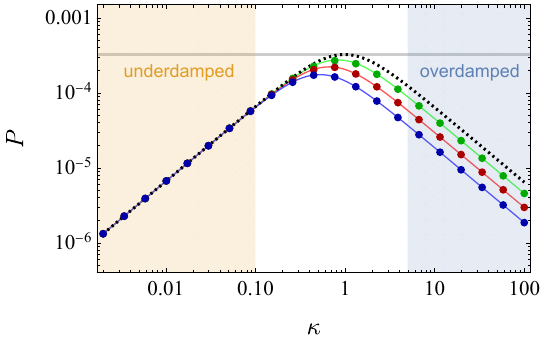}

\caption{Output power of the optimal cycles for different friction coefficients.
We set the control constraints $T_{L}=0.9$, $T_{H}=1$, $\lambda_{H}=1$,
and choose different $\lambda_{L}=1/8$ (blue), $1/4$ (red), $1/2$
(green). The dots and the solid curves show the results of the exact
evolution {[}Eq. (\ref{eq:first-order_eq}){]} and the approximate
evolution {[}Eq. (\ref{eq:dot_V}){]}, respectively. The black dotted
curve shows the maximum output power of sliced cycles, and the gray
horizontal line shows the maximum output power. \label{fig:Bound-of-power}}
\end{figure}

Figure \ref{fig:Bound-of-power} illustrates the output power $P$
of maximum-$\mathcal{H}$ cycles as a function of the friction coefficient
$\kappa$. We set the control constraints $T_{H}=1$, $T_{L}=0.9$,
$\lambda_{H}=1$, and choose different $\lambda_{L}=1/8$ (blue),
$1/4$ (red), $1/2$ (green). The output power of the exact evolution
(dots) agrees well with that of the approximate evolution (solid curves),
and is bounded by the maximum output power of sliced cycles (black
dotted curve) \citep{SupplementaryMaterials}. The maximum output
power in the generic damped situation also recovers that in both the
overdamped and the underdamped limits. The gray horizontal line shows
the maximum output power \citep{SupplementaryMaterials}. The maximum
power can be achieved at the intermediate friction coefficient. This
fact can be explained by the relaxation rate of position $\Delta=\mathrm{Re}(\kappa-\sqrt{\kappa^{2}-4\lambda})$.
Rapid relaxation of position is achieved for large $\Delta$, which
corresponds to $\kappa\sim2\sqrt{\lambda}$.

\textit{Conclusion.} The maximum output power of Brownian heat engines
in the overdamped and the underdamped limits has been obtained previously
\citep{Schmiedl2007,schmiedl2007efficiency,Dechant2017,Chen2022},
but what are the maximum output power and the corresponding optimal
control protocol in generic damped situation (for an arbitrary friction
coefficient) are outstanding unsolved problems in the studies of heat
engines. Previous methods fail \citep{Dechant2017} in this situation
due to the lack of timescale separation.

In this work, we employ the optimal control theory to stochastic thermodynamics,
and solve the outstanding problem of determining the optimal friction
coefficient and the optimal control protocol to achieve the maximum
output power of Brownian heat engines. We construct the optimal cycles
in generic damped situation based on the approximate equation of motion.
These optimal cycles include two isothermal processes, two adiabatic
processes, and an extra isochoric relaxation process at the high stiffness
constraint. It is found that powerful heat engines correspond to the
intermediate friction coefficient, neither too small (underdamped
limit) nor too large (overdamped limit). This argument should also
be valid for quantum heat engines \citep{Uzdin2016,GelbwaserKlimovsky2015,PerarnauLlobet2018}.

Our findings show the effectiveness of the optimal control theory
in stochastic thermodynamics, and provide practical guidance for the
experimental design of Brownian heat engines. It is interesting to
explore the trade-off relation between efficiency and output power
of Brownian heat engines in generic damped situation and also the
optimal control protocol of cycles to achieve the maximum output power
at a given efficiency \citep{Chen1989,shiraishi2016universal,Holubec2016,pietzonka2018universal,Ma2018},
and we leave these problems for future investigation.

\textit{Acknowledgment.} This work is supported by the National Natural
Science Foundation of China (NSFC) under Grants No. 12147157, No.
11775001, No. 11825501, and No. 12375028.

\bibliographystyle{apsrev4-1_addtitle}
\bibliography{power_bound_ref}

\end{document}


\title{Supplementary materials: Optimal control theory for maximum power
of Brownian heat engines}
\author{Jin-Fu Chen}
\thanks{chenjinfu@pku.edu.cn}
\affiliation{School of Physics, Peking University, Beijing, 100871, China}
\author{H. T. Quan}
\thanks{htquan@pku.edu.cn}
\affiliation{School of Physics, Peking University, Beijing, 100871, China}
\affiliation{Collaborative Innovation Center of Quantum Matter, Beijing 100871,
China}
\affiliation{Frontiers Science Center for Nano-optoelectronics, Peking University,
Beijing, 100871, China}
\date{\today}

\maketitle
The supplementary materials are devoted to providing detailed derivations
in the main context. In Sec. \ref{sec:Optimization-in-generic}, we
show the derivations to the optimal control (Eqs. (13)-(17) in the
main text) and the construction of the heat-engine cycles (Fig. 2
of the main text) based on the Pontryagin maximum principle for a
Brownian heat engine in generic damped situation. The derivations
in the underdamped and the overdamped limits are left in Sec. \ref{sec:Optimization-in-underdamped}
and Sec. \ref{sec:Optimization-in-overdamped}, respectively. In Sec.
\ref{sec:Output-power-of}, we discuss the results for the sliced
cycles, which theoretically lead to the bound on the maximum power,
i.e., the black dotted curve in Fig. 3 of the main text.

\tableofcontents{}

\section{Optimization in generic damped situation\label{sec:Optimization-in-generic}}

\renewcommand\theequation{S\arabic{equation}} \renewcommand\thefigure{S\arabic{figure}} \renewcommand\thetable{S\arabic{table}}
\renewcommand{\bibnumfmt}[1]{[S#1]}

We hope to find the optimal control for the following approximate
dynamics

\begin{equation}
\dot{V}=\frac{\kappa^{2}+\lambda}{\kappa^{2}+2\lambda}\alpha V-\frac{\kappa\lambda(2V-T_{b})}{(\kappa^{2}+2\lambda)}.\label{eq:dotV}
\end{equation}
where $\alpha=\dot{\lambda}/\lambda$ is the rate of varying the stiffness.
The work rate during the isothermal process is
\begin{equation}
\dot{W}=\alpha V.
\end{equation}

To maximize the output work 
\begin{equation}
W_{\mathrm{out}}=-W=\int_{0}^{\tau}\alpha Vdt,
\end{equation}
we introduce the Lagrange multipliers $\psi_{V}$ and $\psi_{\lambda}$
and construct the Lagrangian

\begin{equation}
\mathcal{L}=-\alpha V+\psi_{V}\left[\frac{\kappa^{2}+\lambda}{\kappa^{2}+2\lambda}\alpha V-\frac{\kappa\lambda(2V-T_{b})}{\kappa^{2}+2\lambda}-\dot{V}\right]+\psi_{\lambda}(\alpha\lambda-\dot{\lambda}).
\end{equation}
When the control parameters are varied under control constraints,
it is more convenient to define the pseudo-Hamiltonian $\mathcal{H}=\mathcal{H}(V,\lambda,\psi_{V},\psi_{\lambda},\alpha,T_{b})$
as \citep{pontryagin1962}
\begin{align}
\mathcal{H} & =\psi_{V}\dot{V}+\psi_{\lambda}\dot{\lambda}+\mathcal{L}\\
 & =-\alpha V+\psi_{V}\left[\frac{\kappa^{2}+\lambda}{\kappa^{2}+2\lambda}\alpha V-\frac{\kappa\lambda(2V-T_{b})}{\kappa^{2}+2\lambda}\right]+\psi_{\lambda}\lambda\alpha,\label{eq:pseudo_Hamiltonian}
\end{align}
where the state variables are $V$ and $\lambda$, the corresponding
costate variables are $\psi_{V}$ and $\psi_{\lambda}$, and the control
parameters are the rate of stiffness change $\alpha$ and the bath
temperature $T_{b}$. The evolution of the state variables is generated
by

\begin{align}
\dot{V} & =\frac{\partial\mathcal{H}}{\partial\psi_{V}}=\frac{\kappa^{2}+\lambda}{\kappa^{2}+2\lambda}\alpha V-\frac{\kappa\lambda(2V-T_{b})}{\kappa^{2}+2\lambda},\label{eq:dotV_appro}\\
\dot{\lambda} & =\frac{\partial\mathcal{H}}{\partial\psi_{\lambda}}=\alpha\lambda.\label{eq:dotLambda}
\end{align}
The evolution of the costate variables is generated by

\begin{align}
\dot{\psi}_{V} & =-\frac{\partial\mathcal{H}}{\partial V}=\alpha\left[1-\frac{\left(\kappa^{2}+\lambda\right)\psi_{V}}{\kappa^{2}+2\lambda}\right]+\frac{2\kappa\lambda\psi_{V}}{\kappa^{2}+2\lambda},\label{eq:psiVdot}\\
\dot{\psi}_{\lambda} & =-\frac{\partial\mathcal{H}}{\partial\lambda}=\alpha\left[\frac{\kappa^{2}V\psi_{V}}{\left(\kappa^{2}+2\lambda\right)^{2}}-\psi_{\lambda}\right]-\frac{\kappa^{3}(T_{b}-2V)\psi_{V}}{\left(\kappa^{2}+2\lambda\right)^{2}}.\label{eq:psilambdadot}
\end{align}
The control parameters are varied to possibly maximize the pseudo-Hamiltonian
$\mathcal{H}$. We compute its derivatives on the control parameters.
The control of bath temperature is determined by

\begin{equation}
\frac{\partial\mathcal{H}}{\partial T_{b}}=\frac{\kappa\lambda\psi_{V}}{\kappa^{2}+2\lambda},
\end{equation}
which corresponds to a bang-bang control $T_{b}=T_{H}$ for $\psi_{V}>0$
and $T_{L}$ for $\psi_{V}<0$. The control of the stiffness change
rate $\alpha$ is determined from 

\begin{equation}
\frac{\partial\mathcal{H}}{\partial\alpha}=\lambda\psi_{\lambda}+\frac{V\left(\kappa^{2}+\lambda\right)\psi_{V}}{\kappa^{2}+2\lambda}-V,\label{eq:optimalcontrol}
\end{equation}
which does not depend on $\alpha$. The control of the rate of stiffness
change corresponds to a singular control.

\subsection{Isothermal processes}

We solve the singular control in the following. The pseudo-Hamiltonian
$\mathcal{H}$ remains constant during the singular control

\begin{equation}
\frac{d}{dt}\mathcal{H}(V,\lambda,\psi_{V},\psi_{\lambda},\alpha,T_{b})=\dot{V}\frac{\partial\mathcal{H}}{\partial V}+\dot{\lambda}\frac{\partial\mathcal{H}}{\partial\lambda}+\dot{\psi}_{V}\frac{\partial\mathcal{H}}{\partial\psi_{V}}+\dot{\psi}_{\lambda}\frac{\partial\mathcal{H}}{\partial\psi_{\lambda}}+\dot{\alpha}\frac{\partial\mathcal{H}}{\partial\alpha}+\dot{T}_{b}\frac{\partial\mathcal{H}}{\partial T_{b}}.
\end{equation}
The first four terms diminish according to the canonical evolution
equations of the state variables (Eqs. (\ref{eq:dotV_appro}) and
(\ref{eq:dotLambda})) and the costate variables (Eqs. (\ref{eq:psiVdot})
and (\ref{eq:psilambdadot})). The last term is also zero since the
control of bath temperature is bang-bang. Thus, the singular control
requires that
\begin{equation}
\frac{\partial\mathcal{H}}{\partial\alpha}=\lambda\psi_{\lambda}+\frac{V\left(\kappa^{2}+\lambda\right)\psi_{V}}{\kappa^{2}+2\lambda}-V=0.\label{eq:eq1}
\end{equation}
This quantity remains to be zero during the singular control, therefore,
\begin{equation}
\frac{d}{dt}\left(\lambda\psi_{\lambda}+\frac{V\left(\kappa^{2}+\lambda\right)\psi_{V}}{\kappa^{2}+2\lambda}-V\right)=0.\label{eq:differential_eq1}
\end{equation}
Plugging Eqs. (\ref{eq:psiVdot}) and (\ref{eq:psilambdadot}) into
Eq. (\ref{eq:differential_eq1}), we obtain

\begin{equation}
\psi_{V}\left(\lambda T_{b}+2\kappa^{2}V\right)+\left(\kappa^{2}+2\lambda\right)\left(2V-T_{b}\right)=0.\label{eq:eq2}
\end{equation}
Similarly, we have

\begin{equation}
\frac{d}{dt}\left[\psi_{V}\left(\lambda T_{b}+2\kappa^{2}V\right)+\left(\kappa^{2}+2\lambda\right)\left(2V-T_{b}\right)\right]=0.\label{eq:differential_eq2}
\end{equation}
Plugging Eqs. (\ref{eq:psiVdot}) and (\ref{eq:psilambdadot}) into
Eq. (\ref{eq:differential_eq2}), we obtain

\begin{equation}
\psi_{V}\left[\lambda T_{b}\left(\alpha\frac{\kappa^{2}-\lambda}{\kappa^{2}+2\lambda}+2\kappa\frac{\kappa^{2}+\lambda}{\kappa^{2}+2\lambda}\right)+2\alpha\kappa^{2}V\frac{\kappa^{2}-2\lambda}{\kappa^{2}+2\lambda}\right]+T_{b}\left[\alpha\left(\lambda-\kappa^{2}\right)+2\kappa\lambda\right]+2V\left[\alpha\left(3\kappa^{2}+\lambda\right)-2\kappa\lambda\right]=0.\label{eq:eq3}
\end{equation}

We solve $\psi_{V}$ and $\psi_{\lambda}$ from Eqs. (\ref{eq:eq1})
and (\ref{eq:eq2}) as

\begin{align}
\psi_{V} & =\frac{\left(\kappa^{2}+2\lambda\right)\left(T_{b}-2V\right)}{\lambda T_{b}+2\kappa^{2}V},\label{eq:psi1}\\
\psi_{\lambda} & =\frac{V\left[2V\left(2\kappa^{2}+\lambda\right)-\kappa^{2}T_{b}\right]}{\lambda\left(\lambda T_{b}+2\kappa^{2}V\right)},\label{eq:psi2}
\end{align}
Together with Eq. (\ref{eq:eq3}), we obtain the solution to the control
parameter

\begin{equation}
\alpha=\frac{\kappa\lambda(2V-T_{b})\left[2\kappa^{2}V+(\kappa^{2}+2\lambda)T_{b}\right]}{V\left[4\kappa^{4}V+6\kappa^{2}\lambda V+(\kappa^{2}+2\lambda)\lambda T_{b}\right]}.\label{eq:alpha}
\end{equation}
The conserved pseudo-Hamiltonian is obtained 

\begin{equation}
\mathcal{H}=\frac{\kappa\lambda\left(T_{b}-2V\right){}^{2}}{\lambda T_{b}+2\kappa^{2}V}.\label{eq:pseudo_h_general}
\end{equation}
We can solve the stiffness as 

\begin{equation}
\lambda=\frac{2\kappa^{2}V\mathcal{H}}{\kappa\left(T_{b}-2V\right){}^{2}-T_{b}\mathcal{H}}.
\end{equation}

Substituting Eq. (\ref{eq:alpha}) into Eqs. (\ref{eq:dotV_appro})
and (\ref{eq:dotLambda}), we obtain 

\begin{align}
\dot{V} & =-\frac{\kappa^{3}\lambda\left(T_{b}-2V\right)^{2}}{2\lambda^{2}T_{b}+\kappa^{2}\lambda\left(T_{b}+6V\right)+4\kappa^{4}V},\label{eq:dot V approximate}\\
\dot{\lambda} & =\frac{\kappa\lambda^{2}(2V-T_{b})\left[2\kappa^{2}V+T_{b}\left(\kappa^{2}+2\lambda\right)\right]}{V\left[4\kappa^{4}V+6\kappa^{2}\lambda V+T_{b}\left(\kappa^{2}+2\lambda\right)\lambda\right]}.\label{eq:dot lambda approximate}
\end{align}
We determine the work in the isothermal process as 

\begin{align}
W_{\mathrm{iso}} & =\int_{\lambda_{i}}^{\lambda_{f}}V\frac{d\lambda}{\lambda}\nonumber \\
 & =\int_{V_{i}}^{V_{f}}\frac{\kappa T_{b}^{2}-\mathcal{H}T_{b}-4\kappa V^{2}}{\kappa\left(T_{b}-2V\right){}^{2}-\mathcal{H}T_{b}}dV\nonumber \\
 & =\left.\left[-\frac{1}{2}T_{b}\ln\left(\kappa\left(T_{b}-2V\right)^{2}-\mathcal{H}T_{b}\right)-\sqrt{\frac{\mathcal{H}T_{b}}{\kappa}}\tanh^{-1}\left(\sqrt{\frac{\kappa}{\mathcal{H}T_{b}}}\left(T_{b}-2V\right)\right)-V\right]\right|_{V=V_{i}}^{V_{f}}.\label{eq:work_isothermal}
\end{align}
The operation time of the isothermal process is

\begin{align}
\tau_{\mathrm{iso}} & =\int_{V_{i}}^{V_{f}}\frac{dV}{\dot{V}}\nonumber \\
 & =-\int_{V_{i}}^{V_{f}}\frac{2\kappa^{2}\left(T_{b}-2V\right){}^{3}-3\kappa\mathcal{H}\left(T_{b}-2V\right){}^{2}+\mathcal{H}^{2}T_{b}}{\kappa\mathcal{H}\left(T_{b}-2V\right)\left(\kappa\left(T_{b}-2V\right){}^{2}-\mathcal{H}T_{b}\right)}dV\nonumber \\
 & =\left.\left[-\frac{\ln\left(\kappa(T_{b}-2V)^{2}-\mathcal{H}T_{b}\right)}{2\kappa}-\frac{\ln\left(2V-T_{b}\right)}{2\kappa}-\sqrt{\frac{T_{b}}{\kappa\mathcal{H}}}\tanh^{-1}\left(\sqrt{\frac{\kappa}{\mathcal{H}T_{b}}}\left(T_{b}-2V\right)\right)-\frac{2V}{\mathcal{H}}\right]\right|_{V=V_{i}}^{V_{f}}.\label{eq:time_isothermal}
\end{align}

\subsection{Switchings: connecting processes}

The Pontryagin maximum principle requires that in the optimal control
the pseudo-Hamiltonian remains constant \citep{pontryagin1962}. To
construct cycles, we need to find some switchings to connect the two
singular isothermal processes at temperatures $T_{b}=T_{L}$ and $T_{H}$.
In the switchings, the pseudo-Hamiltonian $\mathcal{H}$ remains unchanged.
The switchings can be realized by simultaneously changing $T_{b}$
and $\lambda$ instantaneously, where the heat exchanged can be neglected.
The evolution of the state variables and costate variables during
the switchings is reduced to

\begin{align}
\dot{V} & =\frac{\kappa^{2}+\lambda}{\kappa^{2}+2\lambda}\alpha V,\label{eq:dotV_appro-2}\\
\dot{\lambda} & =\alpha\lambda,\label{eq:dotLambda-2}\\
\dot{\psi}_{V} & =\alpha\left[1-\frac{\left(\kappa^{2}+\lambda\right)\psi_{V}}{\kappa^{2}+2\lambda}\right],\\
\dot{\psi}_{\lambda} & =\alpha\left[\frac{\kappa^{2}V\psi_{V}}{\left(\kappa^{2}+2\lambda\right)^{2}}-\psi_{\lambda}\right].
\end{align}
Thus the values of the potential energy $V$ and its costate variable
$\psi_{V}$ at the beginning and the end of the switchings should
satisfy

\begin{align}
\frac{V_{f}\sqrt{\kappa^{2}+2\lambda_{f}}}{\lambda_{i}} & =\frac{V_{i}\sqrt{2\lambda_{i}+\kappa^{2}}}{\lambda_{i}},\label{eq:connecting process V generalunderdamped}\\
\frac{\lambda_{f}(\psi_{Vf}-2)-\kappa^{2}}{\sqrt{\kappa^{2}+2\lambda_{f}}} & =\frac{\lambda_{i}\left(\psi_{Vi}-2\right)-\kappa^{2}}{\sqrt{\kappa^{2}+2\lambda_{i}}}.\label{eq:connecting process psi_V generalunderdamped}
\end{align}
We calculate the work of the connecting process as 

\begin{align}
W_{\mathrm{con}} & =\int_{\lambda_{i}}^{\lambda_{f}}\frac{d\lambda}{\lambda}V\nonumber \\
 & =\frac{V_{i}}{\lambda_{i}}\sqrt{\kappa^{2}+2\lambda_{i}}\int_{\lambda_{i}}^{\lambda_{f}}\frac{d\lambda}{\sqrt{\kappa^{2}+2\lambda}}\nonumber \\
 & =\frac{V_{i}}{\lambda_{i}}\sqrt{\kappa^{2}+2\lambda_{i}}(\sqrt{\kappa^{2}+2\lambda_{f}}-\sqrt{\kappa^{2}+2\lambda_{i}}).\label{eq:work_connecting}
\end{align}
Notice that the pseudo-Hamiltonian $\mathcal{H}$ should be conserved
in the optimal control of the whole cycle. We will use the conservation
of $\mathcal{H}$ to determine the switching. 

\subsection{Construction of heat engine cycles \label{subsec:Construction-of optimal cycle}}

Using the conservation of $\mathcal{H}$ and the control constraints
of the stiffness $\lambda_{L}\le\lambda\le\lambda_{H}$, we find two
switchings that can be realized by using the low and the high stiffness
constraints $\lambda_{L}$ and $\lambda_{H}$ respectively. We first
consider the switching by increasing $\lambda$ to $\lambda_{H}$.
This happens during the cold isothermal process by simultaneously
changing the stiffness from $\lambda_{2}$ to $\lambda_{H}$ and the
bath temperature from $T_{L}$ to $T_{H}$. The conservation of $\mathcal{H}$
leads to 
\begin{equation}
\mathcal{H}=\frac{\kappa\lambda\left(T_{L}-2V_{2}\right){}^{2}}{\lambda T_{L}+2\kappa^{2}V_{2}}=-\psi_{V3}\frac{\kappa\lambda_{H}(2V_{3}-T_{H})}{\kappa^{2}+2\lambda_{H}},
\end{equation}
where $\psi_{V3}$ is given by Eq. (\ref{eq:psi1}). We solve this
connecting process by using the fact that $V_{3}$ and $\psi_{V3}$
are related to $V_{2}$ and $\psi_{V2}$ according to Eqs. (\ref{eq:connecting process V generalunderdamped})
and (\ref{eq:connecting process psi_V generalunderdamped}).

Similarly, the other switching by decreasing $\lambda$ to $\lambda_{L}$
happens during the hot isothermal process by simultaneously changing
the stiffness from $\lambda_{5}$ to $\lambda_{L}$ and the bath temperature
from $T_{H}$ to $T_{L}$. The conservation of $\mathcal{H}$ leads
to

\begin{equation}
\mathcal{H}=\frac{\kappa\lambda_{5}\left(T_{H}-2V_{5}\right)^{2}}{\lambda_{5}T_{H}+2\kappa^{2}V_{5}}=-\psi_{V6}\frac{\kappa\lambda_{L}(2V_{6}-T_{L})}{\kappa^{2}+2\lambda_{L}}.
\end{equation}
Given specific $\lambda_{L}$, we identify the maximum $\mathcal{H}$
such that point 6 approaches to point 1 on the cold isothermal curve.
When $\mathcal{H}$ is larger than the maximum value, the left switching
connects the two isothermal processes directly without reaching $\lambda_{L}$.

We determine the maximum $\mathcal{H}$ from the conservation of pseudo-Hamiltonian

\begin{equation}
\mathcal{H}=\frac{\kappa\lambda_{5}\left(T_{H}-2V_{5}\right)^{2}}{\lambda_{5}T_{H}+2\kappa^{2}V_{5}}=\frac{\kappa\lambda_{L}\left(T_{L}-2V_{1}\right)^{2}}{\lambda_{L}T_{L}+2\kappa^{2}V_{1}}.
\end{equation}
Points 1 and 5 are on the cold and the hot isothermal curves respectively,
and they satisfy the relation of the switching 

\begin{align}
\frac{V_{1}}{\lambda_{L}}\sqrt{\kappa^{2}+2\lambda_{L}} & =\frac{V_{5}}{\lambda_{5}}\sqrt{\kappa^{2}+2\lambda_{5}},\label{eq:connecting process V generalunderdamped-1}\\
\frac{\lambda_{L}(\psi_{V1}-2)-\kappa^{2}}{\sqrt{\kappa^{2}+2\lambda_{L}}} & =\frac{\lambda_{5}(\psi_{V5}-2)-\kappa^{2}}{\sqrt{\kappa^{2}+2\lambda_{5}}},\label{eq:connecting process psi_V generalunderdamped-1}
\end{align}
and $\psi_{V1}$ and $\psi_{V5}$ are related to $V_{1}$ and $V_{5}$
according to Eq. (\ref{eq:psi1}). Thus the maximum $\mathcal{H}$
is determined by $\lambda_{L}$, $T_{L}$ and $T_{H}$, but we only
obtain the numerical results in generic damped situation. We will
show analytical results of the maximum $\mathcal{H}$ in the underdamped
and the overdamped limits in Sec. \ref{sec:Optimization-in-underdamped}
and Sec. \ref{sec:Optimization-in-overdamped}, respectively. The
output power of the maximum-$\mathcal{H}$ cycle is 
\begin{equation}
P=-\frac{W_{\mathrm{I}}+W_{\mathrm{II}}+W_{\mathrm{IV}}+W_{\mathrm{V}}}{\tau_{\mathrm{I}}+\tau_{\mathrm{III}}+\tau_{\mathrm{IV}}},\label{eq:output_power_expression}
\end{equation}
where we evaluate the work $W_{\mathrm{I}}$ and $W_{\mathrm{IV}}$
of two isothermal processes by Eq. (\ref{eq:work_isothermal}), the
work $W_{\mathrm{II}}$ and $W_{\mathrm{V}}$ of the two connecting
processes by Eq. (\ref{eq:work_connecting}), the duration $\tau_{\mathrm{I}}$
and $\tau_{\mathrm{IV}}$ of two isothermal processes by Eq. (\ref{eq:time_isothermal}),
the duration of the isochoric relaxation by 
\begin{equation}
\tau_{\mathrm{III}}=\frac{\kappa^{2}+2\lambda}{\kappa\lambda}\ln\frac{2V_{3}-T_{H}}{2V_{4}-T_{H}}.\label{eq:time_isochoric}
\end{equation}

\begin{figure}
\includegraphics[width=14cm]{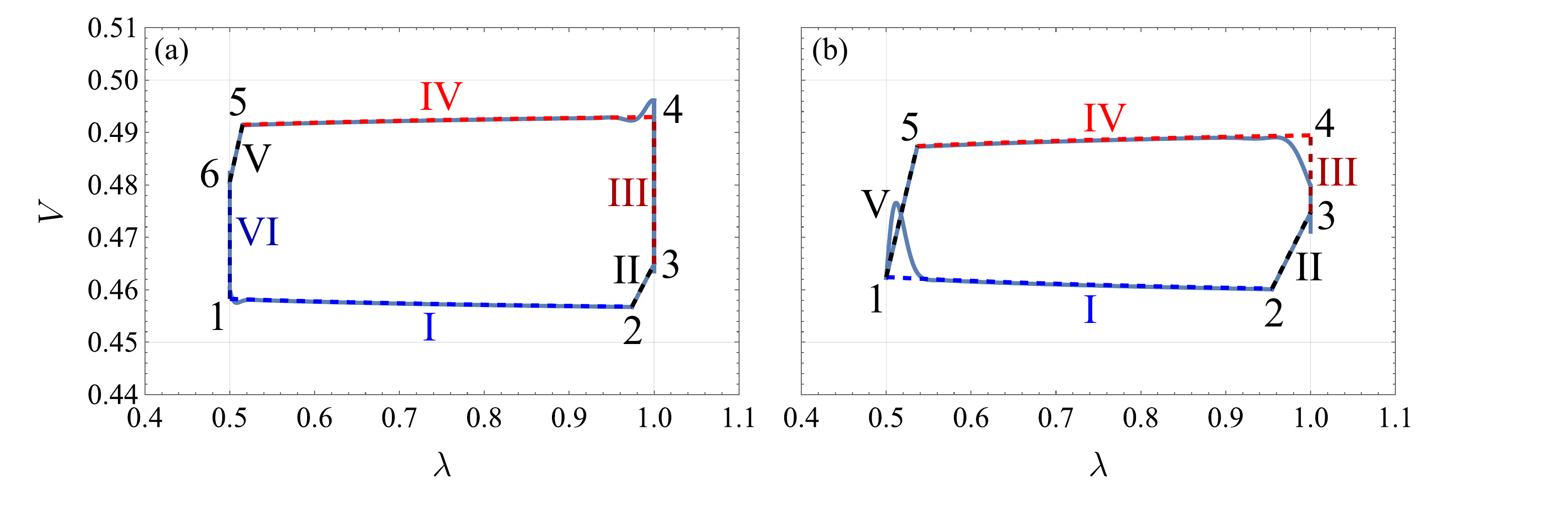}

\caption{Diagrams of heat-engine cycles. (a) $\mathcal{H}=10^{-4}$ (non-maximum-$\mathcal{H}$)
and (b) $\mathcal{H}=2.246\times10^{-4}$ (maximum-$\mathcal{H}$).
The friction coefficient is $\kappa=1$, and the control constraints
$T_{L}=0.9$, $T_{H}=1$, $\lambda_{L}=0.5$ and $\lambda_{H}=1$
are depicted by the horizontal and vertical gray lines. The dashed
curves represent the estimated evolution, while the solid curves are
the exact evolution for the limit cycles.\label{fig:Cycle-diagram general damped}}
\end{figure}

As an illustration, we show cycle diagrams $\lambda$-$V$ in Fig.
\ref{fig:Cycle-diagram general damped} under the parameters $T_{L}=0.9$,
$T_{H}=1$, $\lambda_{L}=0.5$, $\lambda_{H}=1$, and $\kappa=1$.
The control of a whole cycle is determined by $\mathcal{H}$. Increasing
the output power is converted to finding the cycle with larger $\mathcal{H}$.
For larger $\mathcal{H}$, the path of the process VI becomes shorter.
For the maximum $\mathcal{H}$, process VI diminishes and point 6
approaches to point 1, as shown in Fig. \ref{fig:Cycle-diagram general damped}(b).
In the following, we focus on maximum-$\mathcal{H}$ cycles for given
control constraints. To improve the output power, we need to construct
the maximum-$\mathcal{H}$ cycles with appropriate friction coefficient
$\kappa$.

\section{Optimization in underdamped limit \label{sec:Optimization-in-underdamped}}

We discuss the optimization of the cycle in the underdamped limit
$\kappa^{2}\ll\lambda$. The pseudo-Hamiltonian (\ref{eq:pseudo_Hamiltonian})
becomes 
\begin{equation}
\mathcal{H}=-\alpha V+\psi_{V}\left[\frac{1}{2}\alpha V-\kappa(V-\frac{T_{b}}{2})\right]+\psi_{\lambda}\lambda\alpha.\label{eq:pseudo_Hamiltonian-underdamped}
\end{equation}
The evolution of the state variables and costate variables is given
by 

\begin{align}
\dot{V} & =\frac{\partial\mathcal{H}}{\partial\psi_{V}}=\frac{1}{2}\alpha V-\kappa(V-\frac{T_{b}}{2}),\label{eq:dot_V_underdamped}\\
\dot{\lambda} & =\frac{\partial\mathcal{H}}{\partial\psi_{\lambda}}=\alpha\lambda,\\
\dot{\psi}_{V} & =-\frac{\partial\mathcal{H}}{\partial V}=\alpha-\psi_{V}\left(\frac{\alpha}{2}-\kappa\right),\label{eq:dot_psi_V_underdamped}\\
\dot{\psi}_{\lambda} & =-\frac{\partial\mathcal{H}}{\partial\lambda}=-\alpha\psi_{\lambda}.\label{eq:dot_psi_lambda_underdamped}
\end{align}

The control of bath temperature is determined by 

\begin{equation}
\frac{\partial\mathcal{H}}{\partial T_{b}}=\frac{\kappa\psi_{V}}{2},
\end{equation}
which corresponds to a bang-bang control $T_{b}=T_{H}$ for $\psi_{V}>0$
and $T_{L}$ for $\psi_{V}<0$. The control of the stiffness change
rate $\alpha$ is determined from 

\begin{equation}
\frac{\partial\mathcal{H}}{\partial\alpha}=\psi_{\lambda}\lambda+(\frac{\psi_{V}}{2}-1)V.
\end{equation}
We consider the singular control
\begin{equation}
\frac{\partial\mathcal{H}}{\partial\alpha}=\psi_{\lambda}\lambda+(\frac{\psi_{V}}{2}-1)V=0.\label{eq:eq1underdamped}
\end{equation}
This quantity remains to be zero during the singular control, therefore,
\begin{equation}
\frac{d}{dt}\left[\psi_{\lambda}\lambda+(\frac{\psi_{V}}{2}-1)V\right]=0.\label{eq:differential_eq1underdamped}
\end{equation}
Plugging Eqs. (\ref{eq:dot_psi_V_underdamped}) and (\ref{eq:dot_psi_lambda_overdamped})
into Eq. (\ref{eq:differential_eq1underdamped}), we obtain
\begin{equation}
\frac{1}{2}\kappa T_{b}\left(\frac{\psi_{V}}{2}-1\right)+\kappa V=0.\label{eq:eq2underdamped}
\end{equation}
Similarly, we have

\begin{equation}
\frac{d}{dt}\left[\frac{1}{2}\kappa T_{b}\left(\frac{\psi_{V}}{2}-1\right)+\kappa V\right]=0,\label{eq:differential_eq2underdamped}
\end{equation}
which leads to
\begin{equation}
\alpha\left[V-\frac{1}{4}T_{b}\left(\psi_{V}-2\right)\right]+\kappa\left[\frac{1}{2}T_{b}\left(\psi_{V}+2\right)-2V\right]=0.\label{eq:eq3underdamped}
\end{equation}

We solve $\psi_{V}$ and $\psi_{\lambda}$ from Eqs. (\ref{eq:eq1underdamped})
and (\ref{eq:eq2underdamped}) as

\begin{align}
\psi_{V} & =2-\frac{4V}{T_{b}},\label{eq:psi_v_underdamped}\\
\psi_{\lambda} & =\frac{2V^{2}}{\lambda T_{b}}.\label{eq:psi_lambda_underdamped}
\end{align}
We further use Eq. (\ref{eq:eq3underdamped}), and obtain the singular
optimal control satisfying

\begin{equation}
\frac{\dot{\lambda}}{\lambda}=\alpha=\kappa\left(2-\frac{T_{b}}{V}\right).
\end{equation}
Together with the evolution equation (\ref{eq:dot_V_underdamped}),
we find that $V$ and $\alpha$ remains constant in the singular control,
and $\lambda$ is varied in the exponential protocol. Such exponential
protocols have been found as the optimal protocols in the underdamped
limit \citep{Gong2016,Dechant2017,salazar2020work,Chen2022}. 

Substituting Eqs. (\ref{eq:psi_v_underdamped}) and (\ref{eq:psi_lambda_underdamped})
into Eq. (\ref{eq:pseudo_Hamiltonian-underdamped}), the pseudo-Hamiltonian
is explicitly

\begin{equation}
\mathcal{H}=\frac{\kappa\left(T_{b}-2V\right){}^{2}}{T_{b}},\label{eq:pseudo_H_underdamped}
\end{equation}
which is consistent with Eq. (\ref{eq:pseudo_h_general}) by taking
the underdamped limit $\kappa^{2}\ll\lambda$. From Eq. (\ref{eq:pseudo_H_underdamped}),
we obtain the potential energy is constant in the singular control

\begin{equation}
V=\frac{1}{2}(T_{b}\pm\sqrt{\frac{T_{b}\mathcal{H}}{\kappa}}).
\end{equation}
We next consider the switchings between the two isothermal curves
$T_{b}=T_{L}$ and $T_{H}$, and construct the maximum-$\mathcal{H}$
cycle in the underdamped limit.

\subsection*{Construction of heat-engine cycles}

We construct heat-engine cycles in the underdamped limit. In the switchings,
the stiffness is rapidly changed and the heat exchange can be neglected.
The differential equations (\ref{eq:dot_V_underdamped}) and (\ref{eq:dot_psi_V_underdamped})
are reduced to

\begin{align}
\dot{V} & =\frac{1}{2}\alpha V,\label{eq:dotV_underdamped_switching}\\
\dot{\psi}_{V} & =\alpha\left(1-\frac{\psi_{V}}{2}\right).\label{eq:dot_psiV_underdamped_switching}
\end{align}
According to Eq. (\ref{eq:dotV_underdamped_switching}), the work
in a connecting process is obtained as

\begin{equation}
W_{\mathrm{con}}^{\mathrm{over}}=\int_{0}^{\tau_{\mathrm{con}}}\alpha Vdt=2(V_{f}-V_{i}).\label{eq:work_connecting_underdamped}
\end{equation}

We determine the maximum-$\mathcal{H}$ cycle from the conservation
of the pseudo-Hamiltonian in the whole cycle

\begin{equation}
\mathcal{H}=\frac{\kappa\left(T_{H}-2V_{5}\right)^{2}}{T_{H}}=\frac{\kappa\left(T_{L}-2V_{1}\right)^{2}}{T_{L}}.
\end{equation}
Points 1 and 5 are on the cold and the hot isothermal curves respectively.
According to Eqs. (\ref{eq:dotV_underdamped_switching}) and (\ref{eq:dot_psiV_underdamped_switching}),
the two points satisfy the relation of the switching

\begin{align}
\frac{V_{1}}{\sqrt{\lambda_{L}}} & =\frac{V_{5}}{\sqrt{\lambda_{5}}},\label{eq:connecting process V generalunderdamped-1-1}\\
\sqrt{\lambda_{L}}(\psi_{V1}-2) & =\sqrt{\lambda_{5}}(\psi_{V5}-2).\label{eq:connecting process psi_V generalunderdamped-1-1}
\end{align}
We obtain the solution 

\begin{align}
V_{1} & =\frac{1}{4}(T_{L}+\sqrt{T_{H}T_{L}}),\\
V_{5} & =\frac{1}{4}(T_{H}+\sqrt{T_{H}T_{L}}),\\
\mathcal{H} & =\frac{\kappa}{4}(\sqrt{T_{H}}-\sqrt{T_{L}})^{2}.
\end{align}
The rate of the stiffness change is 

\begin{equation}
\alpha=\begin{cases}
2\kappa\frac{\sqrt{T_{H}}-\sqrt{T_{L}}}{\sqrt{T_{H}}+\sqrt{T_{L}}}, & T_{b}=T_{L}\\
-2\kappa\frac{\sqrt{T_{H}}-\sqrt{T_{L}}}{\sqrt{T_{H}}+\sqrt{T_{L}}}. & T_{b}=T_{H}
\end{cases}
\end{equation}
Since $V$ remains constant in the singular control in the underdamped
limit, we obtain $V_{2}=V_{1}$ and $V_{4}=V_{5}$, and the isochoric
process III also diminishes in the maximum-$\mathcal{H}$ cycle, i.e.,
$V_{3}=V_{5}$.

The work and the operation time of each process in the underdamped
limit can be easily calculated. Here we just list the results in the
following. The performed work of each process is 

\begin{align}
W_{\mathrm{I}} & =V_{1}\ln\frac{\lambda_{2}}{\lambda_{L}}=\frac{1}{4}(T_{L}+\sqrt{T_{H}T_{L}})\ln\frac{\lambda_{2}}{\lambda_{L}},\\
W_{\mathrm{II}} & =2V_{3}-2V_{2}=\frac{1}{2}(T_{H}-T_{L}),\\
W_{\mathrm{IV}} & =V_{4}\ln\frac{\lambda_{5}}{\lambda_{H}}=-\frac{1}{4}(T_{H}+\sqrt{T_{H}T_{L}})\ln\frac{\lambda_{2}}{\lambda_{L}},\\
W_{\mathrm{V}} & =2V_{1}-2V_{5}=-\frac{1}{2}(T_{H}-T_{L}),
\end{align}
and the operation time is

\begin{align}
\tau_{\mathrm{I}} & =\tau_{\mathrm{IV}}=\frac{\sqrt{T_{H}}+\sqrt{T_{L}}}{2\kappa(\sqrt{T_{H}}-\sqrt{T_{L}})}\ln\frac{\lambda_{2}}{\lambda_{L}},\\
\tau_{\mathrm{III}} & =0.
\end{align}
According to Eq. (\ref{eq:output_power_expression}), we obtain the
maximum output power in the underdamped limit \citep{Dechant2017,Chen2022}

\begin{equation}
P_{\mathrm{under}}^{*}=\frac{\kappa}{4}(\sqrt{T_{H}}-\sqrt{T_{L}})^{2}.\label{eq:power_underdamped}
\end{equation}
which does not depend on the stiffness constraints. The maximum output
power (\ref{eq:power_underdamped}) in the underdamped limit is proportional
to $\kappa$.

\section{Optimization in overdamped limit \label{sec:Optimization-in-overdamped}}

We discuss the optimization of the cycle in the overdamped limit $\kappa^{2}\gg\lambda$.
The pseudo-Hamiltonian (\ref{eq:pseudo_Hamiltonian}) becomes 

\begin{align}
\mathcal{H} & =-\alpha V+\psi_{V}\left[\alpha V-\frac{\lambda(2V-T_{b})}{\kappa}\right]+\psi_{\lambda}\lambda\alpha.\label{eq:pseudo_Hamiltonian_overdamped}
\end{align}
The evolution of the state variables and costate variables is given
by 

\begin{align}
\dot{V} & =\frac{\partial\mathcal{H}}{\partial\psi_{V}}=\alpha V-\frac{\lambda(2V-T_{b})}{\kappa},\label{eq:dotV_overdamped}\\
\dot{\lambda} & =\frac{\partial\mathcal{H}}{\partial\psi_{\lambda}}=\alpha\lambda,\label{eq:dotLambda_overdamped}\\
\dot{\psi}_{V} & =-\frac{\partial\mathcal{H}}{\partial V}=\psi_{V}\left(\frac{2\lambda}{\kappa}-\alpha\right)+\alpha,\label{eq:dot_psiV_overdamped}\\
\dot{\psi}_{\lambda} & =-\frac{\partial\mathcal{H}}{\partial\lambda}=-\frac{\alpha\kappa\psi_{\lambda}+\psi_{V}\left(T_{b}-2V\right)}{\kappa}.\label{eq:dot_psi_lambda_overdamped}
\end{align}

The control of bath temperature is determined by 

\begin{equation}
\frac{\partial\mathcal{H}}{\partial T_{b}}=\frac{\lambda}{\kappa}\psi_{V},
\end{equation}
which corresponds to a bang-bang control $T_{b}=T_{H}$ for $\psi_{V}>0$
and $T_{L}$ for $\psi_{V}<0$. The control of the stiffness change
rate $\alpha$ is determined from

\begin{equation}
\frac{\partial\mathcal{H}}{\partial\alpha}=\lambda\psi_{\lambda}+V\left(\psi_{V}-1\right).
\end{equation}
We consider the singular control 
\begin{equation}
\frac{\partial\mathcal{H}}{\partial\alpha}=\lambda\psi_{\lambda}+V\left(\psi_{V}-1\right)=0.\label{eq:eq1overdamped}
\end{equation}
This quantity remains to be zero during the singular control, therefore,
\begin{equation}
\frac{d}{dt}\left[\lambda\psi_{\lambda}+V\left(\psi_{V}-1\right)\right]=0.\label{eq:differential_eq1overdamped}
\end{equation}
Plugging Eqs. (\ref{eq:dotV_overdamped}) and (\ref{eq:dotLambda_overdamped})
into Eq. (\ref{eq:differential_eq1overdamped}), we obtain 
\begin{equation}
2V\left(\psi_{V}+1\right)-T_{b}=0\label{eq:eq2overdamped}
\end{equation}
Similarly, we have

\begin{equation}
\frac{d}{dt}\left[2V\left(\psi_{V}+1\right)-T_{b}\right]=0,
\end{equation}
which leads to
\begin{equation}
\alpha\kappa\left[T_{b}-2V\left(\psi_{V}+3\right)\right]-2\lambda T_{b}\left(\psi_{V}+1\right)+4\lambda V=0.\label{eq:eq3overdamped}
\end{equation}

We solve $\psi_{V}$ and $\psi_{\lambda}$ from Eqs. (\ref{eq:eq1overdamped})
and (\ref{eq:eq2overdamped}) as

\begin{align}
\psi_{V} & =\frac{T_{b}}{2V}-1,\label{eq:psi_V_overdamped}\\
\psi_{\lambda} & =\frac{2V}{\lambda}-\frac{T_{b}}{2\lambda},\label{eq:psi_lambda_overdamped}
\end{align}
We further use Eq. (\ref{eq:eq3overdamped}) and obtain the singular
optimal control as

\begin{equation}
\alpha=\frac{\lambda}{\kappa}\left(1-\frac{T_{b}^{2}}{4V^{2}}\right).
\end{equation}
The differential equations for the optimal control are 
\begin{align}
\dot{\lambda} & =\frac{\lambda^{2}}{\kappa}\left(1-\frac{T_{b}^{2}}{4V^{2}}\right),\\
\dot{V} & =-\frac{\lambda}{\kappa V}\left(V-\frac{T_{b}}{2}\right)^{2}.
\end{align}
Under the initial condition $\lambda(0)=\lambda_{0}$ and $V(0)=V_{0}$,
we obtain the solution

\begin{align}
V(t) & =V_{0}-\frac{\lambda_{0}t\left(T_{b}-2V_{0}\right){}^{2}}{4\kappa V_{0}},\\
\lambda(t) & =\frac{\kappa\lambda_{0}\left[4\kappa V_{0}^{2}-\lambda_{0}t\left(T_{b}-2V_{0}\right){}^{2}\right]}{\left[\lambda_{0}t\left(T_{b}-2V_{0}\right)+2\kappa V_{0}\right]^{2}}.
\end{align}
which coincides with the control scheme obtained in Refs. \citep{Schmiedl2007,schmiedl2007efficiency}. 

Substituting Eqs. (\ref{eq:psi_V_overdamped}) and (\ref{eq:psi_lambda_overdamped})
into Eq. (\ref{eq:pseudo_Hamiltonian_overdamped}), the pseudo-Hamiltonian
is explicitly 

\begin{equation}
\mathcal{H}=\frac{\lambda\left(T_{b}-2V\right){}^{2}}{2\kappa V}.
\end{equation}
This expression is consistent with Eq. (\ref{eq:pseudo_h_general})
by taking the overdamped limit $\kappa^{2}\gg\lambda$. Therefore,
the potential energy is related to the stiffness as

\begin{equation}
V=\frac{T_{b}}{2}+\frac{\kappa\mathcal{H}}{4\lambda}\left(1\pm\sqrt{\frac{4\lambda T_{b}}{\kappa\mathcal{H}}+1}\right).
\end{equation}

The work in an isothermal process can be obtained as 

\begin{align}
W_{\mathrm{iso}}^{\mathrm{over}} & =\int_{\lambda_{i}}^{\lambda_{f}}\frac{V}{\lambda}d\lambda\nonumber \\
 & =\int_{V_{i}}^{V_{f}}\frac{T_{b}+2V}{T_{b}-2V}dV\nonumber \\
 & =V_{i}-V_{f}+T_{b}\ln\frac{2V_{i}-T_{b}}{2V_{f}-T_{b}}.\label{eq:work_isothermal_overdamped}
\end{align}
The operation time is 

\begin{align}
\tau_{\mathrm{iso}}^{\mathrm{over}} & =\int_{V_{i}}^{V_{f}}\frac{dV}{\dot{V}}\nonumber \\
 & =-\int_{V_{i}}^{V_{f}}\frac{4\kappa V}{\lambda\left(2V-T_{b}\right)^{2}}dV\nonumber \\
 & =2\frac{V_{i}-V_{f}}{\mathcal{H}}.\label{eq:time_isothermal_overdamped}
\end{align}
Thus, we obtain the work and operation time for processes I and IV
in the overdamped limit.

\subsection*{Construction of heat-engine cycles}

In the switchings, the stiffness is suddenly changed and the heat
exchange can be neglected. The differential equations (\ref{eq:dotV_overdamped})
and (\ref{eq:dot_psiV_overdamped}) are reduced to

\begin{align}
\dot{V} & =\alpha V,\label{eq:dotV_overdamped_switching}\\
\dot{\psi}_{V} & =(1-\psi_{V})\alpha.\label{eq:dotpsiV_overdamped_switching}
\end{align}
According to Eq. (\ref{eq:dotV_overdamped_switching}), the work in
a connecting process is obtained as

\begin{equation}
W_{\mathrm{con}}^{\mathrm{over}}=\int_{0}^{\tau_{\mathrm{con}}}\alpha Vdt=V_{f}-V_{i}.\label{eq:work_connecting_overdamped}
\end{equation}
When the stiffness is switched to $\lambda_{H}$, an isochoric relaxation
process follows

\begin{align}
\dot{V} & =-\frac{\lambda_{H}(2V-T_{H})}{\kappa},\label{eq:dotV_overdamped_isochoric}\\
\dot{\psi}_{V} & =\psi_{V}\frac{2\lambda_{H}}{\kappa}.\label{eq:dotpsiV_overdamped_isochoric}
\end{align}

We first calculate the left switching of the maximum-$\mathcal{H}$
cycle in the overdamped limit. According to Eqs. (\ref{eq:dotV_overdamped_switching})
and (\ref{eq:dotpsiV_overdamped_switching}), the two boundary points
1 and 5 satisfy

\begin{align}
\frac{V_{5}}{\lambda_{5}} & =\frac{V_{1}}{\lambda_{L}},\label{eq:left_connecting_process_overdamped}\\
(1-\psi_{V5})\lambda_{5} & =(1-\psi_{V1})\lambda_{L},
\end{align}
where $\psi_{V1}$ and $\psi_{V5}$ are related to $V_{1}$ and $V_{5}$
by Eq. (\ref{eq:psi_V_overdamped}). The two points are located on
two isothermal curves, and the conservation of $\mathcal{H}$ leads
to 
\begin{equation}
\mathcal{H}=\frac{\lambda_{L}\left(T_{L}-2V_{1}\right){}^{2}}{2\kappa V_{1}}=\frac{\lambda_{5}(T_{H}-2V_{5})^{2}}{2\kappa V_{5}}.\label{eq:left_connecting_process_overdamped_conservation_H}
\end{equation}
From Eqs. (\ref{eq:left_connecting_process_overdamped})-(\ref{eq:left_connecting_process_overdamped_conservation_H}),
we obtain the solution of the left switching for the maximum-$\mathcal{H}$
cycle

\begin{align}
\mathcal{H} & =\frac{(T_{H}-T_{L})^{2}\lambda_{L}}{4(T_{H}+3T_{L})\kappa},\\
V_{1} & =\frac{1}{8}(T_{H}+3T_{L}),\label{eq:V1_overdamped}\\
\psi_{V1} & =-\frac{T_{H}-T_{L}}{T_{H}+3T_{L}},\\
\lambda_{5} & =\frac{3T_{H}+T_{L}}{T_{H}+3T_{L}}\lambda_{L},\\
V_{5} & =\frac{1}{8}(3T_{H}+T_{L}),\label{eq:V5_overdamped}\\
\psi_{V5} & =\frac{T_{H}-T_{L}}{3T_{H}+T_{L}}.
\end{align}
To achieve large $\mathcal{H}$, we can possibly choose large $\lambda_{L}$,
but $\lambda_{5}$ should be less than $\lambda_{H}$, which gives

\begin{equation}
\lambda_{L}\le\frac{T_{H}+3T_{L}}{3T_{H}+T_{L}}\lambda_{H},
\end{equation}
and the pseudo-Hamiltonian is less than 

\begin{equation}
\mathcal{H}\le\frac{(T_{H}-T_{L})^{2}\lambda_{H}}{4(3T_{H}+T_{L})\kappa}.\label{eq:H_bound_approximate}
\end{equation}

We next solve the right switching with given $\mathcal{H}$. The potential
energy and the stiffness for the points 2 and 3 satisfy

\begin{align}
\mathcal{H} & =\frac{\lambda_{2}\left(T_{L}-2V_{2}\right){}^{2}}{2\kappa V_{2}}=-\psi_{V3}\frac{\lambda_{H}}{\kappa}(2V_{3}-T_{H}),\label{eq:overdamped_right_quench_eq1}\\
\frac{V_{3}}{\lambda_{H}} & =\frac{V_{2}}{\lambda_{2}},\\
(1-\psi_{V3})\lambda_{H} & =(1-\psi_{V2})\lambda_{2},\label{eq:overdamped_right_quench_eq3}\\
\psi_{V2} & =\frac{T_{L}}{2V_{2}}-1.\label{eq:overdamped_right_quench_eq4}
\end{align}
We eliminate $V_{3}$ and $\psi_{V3}$ and obtain 

\begin{align}
\mathcal{H} & =\left[(2-\frac{T_{L}}{2V_{2}})\frac{\lambda_{2}}{\lambda_{H}}-1\right]\frac{\lambda_{H}}{\kappa}(2\frac{\lambda_{H}V_{2}}{\lambda_{2}}-T_{H}),\label{eq:H_over_Eq1}\\
V_{2} & =\frac{T_{L}}{2}+\frac{\kappa\mathcal{H}}{4\lambda_{2}}\left(1-\sqrt{\frac{4\lambda T_{L}}{\kappa\mathcal{H}}+1}\right).\label{eq:V2_over_eq2}
\end{align}
We do not find simple algebra expressions of $V_{2}$ and $\lambda_{2}$.
However, Eqs. (\ref{eq:H_over_Eq1}) and (\ref{eq:V2_over_eq2}) have
real solutions only when $\mathcal{H}$ is less than a maximum value
$\mathcal{H}_{\mathrm{max}}^{\mathrm{over}}$. To evaluate $\mathcal{H}_{\mathrm{max}}^{\mathrm{over}}$,
we first eliminate $\lambda_{2}$ and obtain the solution $V_{2}=T_{H}\chi$,
where $\chi>T_{L}/(2T_{H})$ is the root of a fourth-order polynomial
\begin{equation}
f(\chi,\frac{\kappa\mathcal{H}}{\lambda_{H}T_{H}})=0,
\end{equation}
where the polynomial is explicitly

\begin{equation}
f(\chi,\phi)=\phi^{2}\left[\left(\theta-4\chi\right)-\left(\theta-2\chi\right){}^{2}\right]+\phi\left(\theta-2\chi\right){}^{2}\left(1-\theta+4\chi\right)-\left(\theta-2\chi\right){}^{4}.
\end{equation}
We use $\theta=T_{L}/T_{H}$ to denote the temperature ratio. If the
polynomial has a real root satisfying $\chi>T_{L}/(2T_{H})$ , the
maximum value $f(\chi^{*},\phi)$ on $\chi>T_{L}/(2T_{H})$ should
be positive. Thus, the maximum point satisfies

\begin{align}
f(\chi^{*},\phi) & \ge0,\\
\left.\frac{\partial f}{\partial\chi}\right|_{\chi=\chi^{*}} & =0.
\end{align}
Then we solve the above formulas with the equality and obtain $\phi(\theta)$
as a positive root of a third-order polynomial

\begin{equation}
\phi^{3}+(11-3\theta)\phi^{2}+\left(3\theta^{2}+14\theta-1\right)\phi-(1-\theta)^{2}\theta=0.\label{eq:third-order polynomial}
\end{equation}
The positive root is explicitly

\begin{equation}
\phi(\theta)=\frac{1}{3}\sqrt[3]{\Upsilon(\theta)}+\frac{4}{3}\frac{(31-27\theta)}{\sqrt[3]{\Upsilon(\theta)}}+\theta-\frac{11}{3},\label{eq:root_third-order polynomial}
\end{equation}
where

\begin{equation}
\Upsilon(\theta)=-\frac{729\theta^{2}}{2}+1809\theta-\frac{2761}{2}+\frac{3}{2}i\sqrt{3(1-\theta)(27\theta+5)^{3}}.
\end{equation}
The maximum value $\mathcal{H}_{\mathrm{over}}^{\mathrm{max}}$ is
represented as

\begin{equation}
\mathcal{H}_{\mathrm{over}}^{\mathrm{max}}=\frac{\lambda_{H}T_{H}}{\kappa}\phi(\frac{T_{L}}{T_{H}}).
\end{equation}

According to Eqs. (\ref{eq:work_isothermal_overdamped}) and (\ref{eq:work_connecting_overdamped}),
we obtain the expression of the performed work for each process as 

\begin{align}
W_{\mathrm{I}} & =V_{1}-V_{2}+T_{L}\ln\frac{2V_{1}-T_{L}}{2V_{2}-T_{L}},\\
W_{\mathrm{II}} & =V_{3}-V_{2},\\
W_{\mathrm{IV}} & =V_{4}-V_{5}+T_{H}\ln\frac{2V_{4}-T_{H}}{2V_{5}-T_{H}},\\
W_{V} & =V_{1}-V_{5},
\end{align}
and the operation time according to Eq. (\ref{eq:time_isothermal_overdamped})
is

\begin{align}
\tau_{\mathrm{I}} & =\frac{2(V_{1}-V_{2})}{\mathcal{H}},\\
\tau_{\mathrm{III}} & =\frac{\kappa}{2\lambda_{H}}\ln\frac{2V_{3}-T_{H}}{2V_{4}-T_{H}},\\
\tau_{\mathrm{IV}} & =\frac{2(V_{4}-V_{5})}{\mathcal{H}}.
\end{align}
The output power of the maximum-$\mathcal{H}$ cycle according to
Eq. (\ref{eq:output_power_expression}) is

\begin{equation}
P_{\mathrm{over}}=-\frac{2V_{1}-2V_{2}+V_{3}+V_{4}-2V_{5}+T_{L}\ln\frac{2V_{1}-T_{L}}{2V_{2}-T_{L}}+T_{H}\ln\frac{2V_{4}-T_{H}}{2V_{5}-T_{H}}}{\frac{8}{\lambda_{L}}\frac{(T_{H}+3T_{L})}{(T_{H}-T_{L})^{2}}(V_{1}-V_{2}+V_{4}-V_{5})+\frac{1}{2\lambda_{H}}\ln\frac{2V_{3}-T_{H}}{2V_{4}-T_{H}}}\times\frac{1}{\kappa}.\label{eq:maximum_output_power_overdamped}
\end{equation}
We notice that the potential energies $V_{j}$ of the maximum-$\mathcal{H}$
cycle are solved previously by Eqs. (\ref{eq:V1_overdamped}), (\ref{eq:V5_overdamped}),
and (\ref{eq:overdamped_right_quench_eq1})-(\ref{eq:overdamped_right_quench_eq4}),
and are independent of the friction coefficient $\kappa$. Therefore,
we conclude the maximum output power (\ref{eq:maximum_output_power_overdamped})
in the overdamped limit is inversely proportional to $\kappa$.

\begin{figure}
\includegraphics[width=7cm]{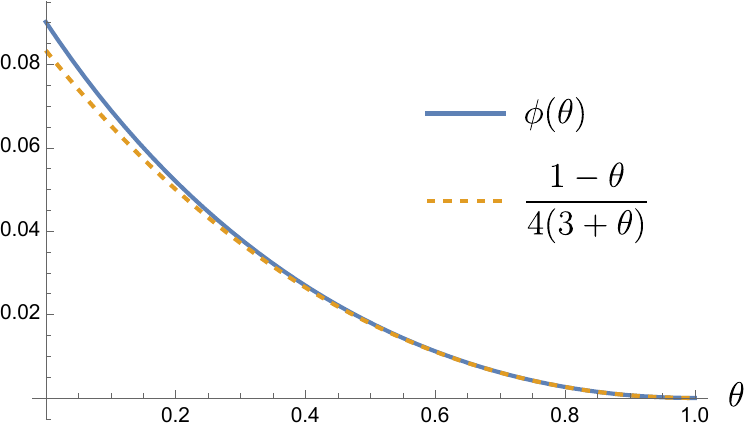}

\caption{Comparison of the precise and approximate results of the maximum power
(Eq. (\ref{eq:poverboundexact})) in the overdamped limit. \label{fig:Comparison-of-Eqs.}}
\end{figure}

To obtain the maximum power in the overdamped limit, we consider the
sliced cycle constructed around the above two switchings. This follows
from a simple argument that one can always divide a cycle into several
slices, and some slice has the largest power among all slices. For
the sliced cycles, the output power becomes equal to the pseudo-Hamiltonian.
We will show this in Sec. \ref{sec:Output-power-of}. Therefore, the
maximum power in the overdamped limit according to Eq. (\ref{eq:third-order polynomial})
is \textbf{
\begin{equation}
P_{\mathrm{over}}^{*}=\frac{\lambda_{H}T_{H}}{\kappa}\phi(\frac{T_{L}}{T_{H}})\approx\frac{(T_{H}-T_{L})^{2}\lambda_{H}}{4(3T_{H}+T_{L})\kappa}\label{eq:poverboundexact}
\end{equation}
}where $\phi(\theta)$ is given by Eq. (\ref{eq:root_third-order polynomial}).
We compare the precise and approximate results of the maximum power
in Fig. \ref{fig:Comparison-of-Eqs.}.

\section{Output power of sliced cycle\label{sec:Output-power-of}}

\subsection{Generic damped situation}

\begin{figure}
\includegraphics[width=7cm]{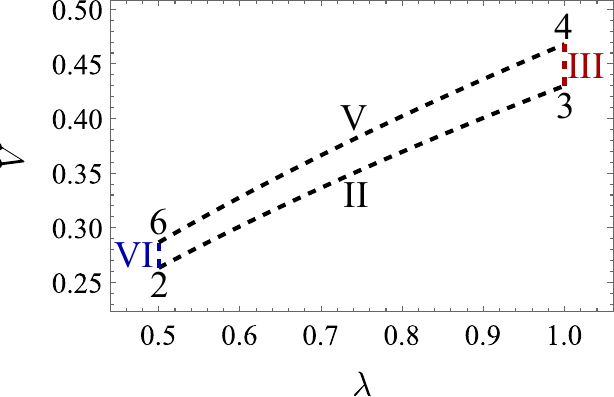}

\caption{Diagram of a sliced cycle. It is an Otto cycle consisting of two adiabatic
processes and two isochoric relaxation processes. \label{fig:Diagram-of-a}}
\end{figure}

We next consider the sliced cycles in generic damped situation. The
diagram of a sliced cycle is shown in Fig. \ref{fig:Diagram-of-a}.
Since the operation time of the two relaxation processes is short
enough, we can neglect the change of stiffness during the two processes
and just consider them as isochoric processes. Thus the cycle becomes
an Otto cycle.

The potential energy of point 4 is related to that of point 3 as 
\begin{equation}
V_{4}=V_{3}+dV_{3}.
\end{equation}
According to the switchings, we obtain the potential energy at points
2 and 6 as

\begin{align}
\frac{V_{2}}{\lambda_{2}}\sqrt{\kappa^{2}+2\lambda_{2}} & =\frac{V_{3}}{\lambda_{H}}\sqrt{\kappa^{2}+2\lambda_{H}},\\
\frac{V_{6}}{\lambda_{2}}\sqrt{\kappa^{2}+2\lambda_{2}} & =\frac{V_{4}}{\lambda_{H}}\sqrt{\kappa^{2}+2\lambda_{H}}.
\end{align}
The work in the connecting processes is 
\begin{align}
W_{\mathrm{II}} & =(\sqrt{\kappa^{2}+2\lambda_{H}}-\sqrt{\kappa^{2}+2\lambda_{2}})\frac{V_{3}}{\lambda_{H}}\sqrt{\kappa^{2}+2\lambda_{H}},\\
W_{\mathrm{V}} & =(\sqrt{\kappa^{2}+2\lambda_{2}}-\sqrt{\kappa^{2}+2\lambda_{H}})\frac{V_{4}}{\lambda_{H}}\sqrt{\kappa^{2}+2\lambda_{H}}.
\end{align}
The operation time of the isochoric processes is 
\begin{align}
\tau_{\mathrm{III}} & =\frac{\kappa^{2}+2\lambda_{H}}{2\kappa\lambda_{H}}\ln\frac{T_{H}-2V_{3}}{T_{H}-2V_{4}}\nonumber \\
 & =\frac{\kappa^{2}+2\lambda_{H}}{\kappa\lambda_{H}}\frac{dV_{3}}{T_{H}-2V_{3}},
\end{align}
and

\begin{align}
\tau_{\mathrm{VI}} & =\frac{\kappa^{2}+2\lambda_{2}}{2\kappa\lambda_{2}}\ln\frac{2V_{6}-T_{L}}{2V_{2}-T_{L}}\nonumber \\
 & =\frac{\kappa^{2}+2\lambda_{2}}{\kappa\lambda_{2}}\frac{dV_{2}}{2V_{2}-T_{L}}.
\end{align}
The output power is 
\begin{align}
P^{\mathrm{slice}} & =-\frac{W_{\mathrm{II}}+W_{\mathrm{V}}}{\tau_{\mathrm{III}}+\tau_{\mathrm{VI}}}\nonumber \\
 & =\sqrt{\lambda_{H}}T_{H}\frac{\frac{\kappa}{\sqrt{\lambda_{H}}}(1-\sqrt{\frac{\kappa^{2}+2\lambda_{2}}{\kappa^{2}+2\lambda_{H}}})}{\frac{T_{H}}{T_{H}-2V_{3}}+\frac{\kappa^{2}+2\lambda_{2}}{\kappa^{2}+2\lambda_{H}}\frac{T_{H}}{2V_{3}\frac{\lambda_{2}}{\lambda_{H}}-T_{L}\sqrt{\frac{\kappa^{2}+2\lambda_{2}}{\kappa^{2}+2\lambda_{H}}}}}.\label{eq:power_slice}
\end{align}
Let us define some dimensionless quantities 
\begin{align}
u & =\frac{\kappa}{\sqrt{\lambda_{H}}},\\
y & =\frac{\lambda_{2}}{\lambda_{H}},\\
z & =\frac{2V_{3}}{T_{H}},\\
v & =\sqrt{\frac{u^{2}+2y}{u^{2}+2}}.
\end{align}
Then we rewrite the output power as 

\begin{align}
P^{\mathrm{slice}} & =\sqrt{\lambda_{H}}T_{H}h(z,v,y),
\end{align}
where we have defined a function
\begin{equation}
h(z,v,y)\equiv\frac{(1-v)}{\frac{1}{1-z}+\frac{v^{2}}{zy-\theta v}}\sqrt{\frac{2(v^{2}-y)}{1-v^{2}}}.\label{eq:h_generaldamped}
\end{equation}

We optimize the function $h(z,v,y)$ under the constraints $v\theta/z<y<v^{2}<1$,
$0<z<1$ to obtain a bound on the output power. We first choose the
optimal 
\begin{equation}
z^{*}=\frac{v\left(\theta+\sqrt{y}\right)}{v\sqrt{y}+y},
\end{equation}
and we obtain

\begin{equation}
h(z^{*},v,y)=\frac{\sqrt{2}(1-v)(y-\theta v)}{\left(v+\sqrt{y}\right)^{2}}\sqrt{\frac{v^{2}-y}{1-v^{2}}}.
\end{equation}
We further optimize $v$ and $y$ to obtain the global upper bound
$h(z,v,y)\leq h(z^{*},v^{*},y^{*})\equiv h_{\mathrm{bound}}$, which
can be represented as the positive root $x$ of the following polynomial

\begin{align}
 & 729(\theta+1)^{2}x^{12}-108\left(\theta^{4}+690\theta^{3}+7893\theta^{2}+12888\theta+5940\right)x^{10}\nonumber \\
 & +4\left(\theta^{6}+2610\theta^{5}+574425\theta^{4}-3453408\theta^{3}-148824\theta^{2}-2259360\theta-1219536\right)x^{8}\nonumber \\
 & -128\left(3\theta^{7}+2131\theta^{6}+85745\theta^{5}-385087\theta^{4}+983572\theta^{3}-81252\theta^{2}+48816\theta+11880\right)x^{6}\nonumber \\
 & +512\left(19\theta^{8}+1288\theta^{7}+31966\theta^{6}-82560\theta^{5}-157805\theta^{4}-46328\theta^{3}+11892\theta^{2}-16\theta+8\right)x^{4}\nonumber \\
 & -8192\theta^{2}\left(3\theta^{7}+39\theta^{6}+1294\theta^{5}+3180\theta^{4}+2883\theta^{3}+811\theta^{2}-20\theta+2\right)x^{2}+16384(\theta-1)^{4}\theta^{4}(\theta+1)^{2}=0,\label{eq:root to give h_bound}
\end{align}
where $v^{*}$ and $y^{*}$ are solved from 

\begin{align}
-2\theta v^{2}+v\sqrt{y}(\theta-2v)+2vy+y^{3/2} & =0,\\
\theta v^{3}-2\left(v^{2}-1\right)y^{3/2}+2\theta v\left(v^{2}-1\right)\sqrt{y}+y\left(\theta+v^{3}-(\theta+1)v^{2}-(\theta+1)v\right)+y^{2} & =0.
\end{align}
The bound on the output power is expressed by 
\begin{equation}
P_{\mathrm{bound}}=\sqrt{\lambda_{H}}T_{H}h_{\mathrm{bound}}.
\end{equation}

We can also maximize the power with given $\kappa$, namely, with
given $u$ maximizing
\begin{equation}
h(z^{*},\sqrt{\frac{u^{2}+2y}{u^{2}+2}},y)=\frac{u(1-\sqrt{\frac{u^{2}+2y}{u^{2}+2}})(y-\theta\sqrt{\frac{u^{2}+2y}{u^{2}+2}})}{\left(\sqrt{\frac{u^{2}+2y}{u^{2}+2}}+\sqrt{y}\right)^{2}}.
\end{equation}
The optimal choose of $y^{*}(u)$ with given $u$ is the root of 
\begin{align}
 & 2\theta^{2}u^{2}+y^{3/2}\left(\theta^{2}+4\theta-2\theta u^{2}-2u^{2}\right)+y^{2}\left(-2\theta-2u^{2}\right)\nonumber \\
 & +y\left(3\theta^{2}-2\theta u^{2}+2u^{2}\right)+4\theta u^{2}\sqrt{y}+(-2\theta-4)y^{5/2}+y^{7/2}-y^{3}=0.
\end{align}
We denote the maximum value as 
\begin{equation}
h^{*}\equiv h(z^{*},\sqrt{\frac{u^{2}+2y}{u^{2}+2}},y^{*}(u)).\label{eq:h^*}
\end{equation}
We use the above results Eqs. (\ref{eq:root to give h_bound}) and
(\ref{eq:h^*}) to plot the horizontal gray line and the black dotted
curve in Fig. 3 of the main text.

\subsection{Overdamped limit}

In the overdamped limit $\kappa^{2}\gg\lambda$ or $u\gg1$, the power
Eq. (\ref{eq:power_slice}) becomes 
\begin{align}
P_{\mathrm{over}}^{\mathrm{slice}} & =\sqrt{\lambda_{H}}T_{H}h_{\mathrm{over}}(z,y,u),
\end{align}
where the function $h_{\mathrm{over}}(z,y,u)$ according to Eq. (\ref{eq:h_generaldamped})
is 
\begin{equation}
h_{\mathrm{over}}(z,y,u)=\frac{1-y}{\frac{1}{1-z}+\frac{1}{zy-\theta}}\frac{1}{u}.
\end{equation}
We have used the fact that $1-v\approx(1-y)/u^{2}$. For $0<y<1$
and $\theta/y<z<1$, the maximum of $h_{\mathrm{over}}$ is exactly
given by $\phi(\theta)/u$ solved from Eq. (\ref{eq:third-order polynomial}). 

\subsection{Underdamped limit}

In the underdamped limit $\kappa^{2}\ll\lambda$ or $u\gg1$, the
power Eq. (\ref{eq:power_slice}) becomes 

\begin{align}
P_{\mathrm{under}}^{\mathrm{slice}} & =\sqrt{\lambda_{H}}T_{H}h_{\mathrm{under}}(z,y,u),
\end{align}
where the function $h_{\mathrm{under}}(z,y,u)$ according to Eq. (\ref{eq:h_generaldamped})
is
\begin{equation}
h_{\mathrm{under}}(z,y,u)=\frac{\left(1-\sqrt{y}\right)(1-z)\left(\sqrt{y}z-\theta\right)}{\sqrt{y}-\theta}u.\label{eq:h_underdamped}
\end{equation}
We have used the fact that $v\approx y$. For $\theta^{2}<y<1$ and
$\theta/\sqrt{y}<z<1$, the maximum of $h_{\mathrm{under}}$ is 
\begin{equation}
h_{\mathrm{under}}^{*}=\frac{1}{4}u(1-\sqrt{\theta})^{2},
\end{equation}
which is reached by $y=\theta$ and $z=(1+\sqrt{\theta})/2$. The
output power then coincides with Eq. (\ref{eq:power_underdamped}).

\subsection{Maximum output power in generic damped situation}

\begin{figure}
\includegraphics[width=7cm]{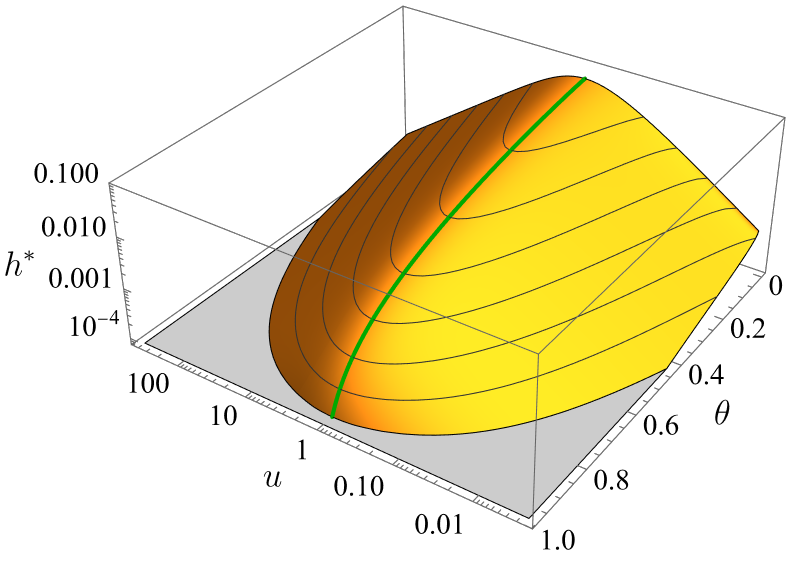}

\caption{The maximum output power $P^{*}=\sqrt{\lambda_{H}}T_{H}h^{*}$ as
a function of the friction coefficient $u$ and the temperature ratio
$\theta$. The green solid curve shows the optimal choice of the friction
coefficient to maximize the output power $P_{\mathrm{bound}}=\sqrt{\lambda_{H}}T_{H}h_{\mathrm{bound}}$.
Here we choose $\lambda_{H}=1$ and $T_{H}=1$. \label{fig:Bound-on-output}}

\end{figure}

We show the maximum output power $P^{*}=\sqrt{\lambda_{H}}T_{H}h^{*}$
as a function of the friction coefficient $u$ and the temperature
ratio $\theta$ in Fig. \ref{fig:Bound-on-output}. For given $\theta$,
large output power can be obtained by choosing an appropriate friction
ratio reflecting by $u$. The bound of the output power $P_{\mathrm{bound}}=\sqrt{\lambda_{H}}T_{H}h_{\mathrm{bound}}$
is shown by the green solid curve. Here we choose $\lambda_{H}=1$
and $T_{H}=1$. The definitions of $h_{\mathrm{bound}}$ and $h^{*}$
are given by Eqs. (\ref{eq:root to give h_bound}) and (\ref{eq:h^*}),
respectively.

\bibliographystyle{apsrev4-1}
\bibliography{ref}